\def\draftversion{false}
\RequirePackage{ifthen}
\ifthenelse{\equal{\draftversion}{true}}{
  \documentclass[aps,prb,10pt,galley,amsmath,amssymb,showpacs,
                 superscriptaddress]{revtex4}
}{
  \documentclass[aps,prb,10pt,twocolumn,showpacs,
                 superscriptaddress]{revtex4-1}
}

\usepackage{times}
\usepackage{amsmath}
\usepackage{amssymb}
\usepackage{graphicx}
\usepackage[usenames, dvipsnames]{color} 
\usepackage{bm} 
\usepackage{subfigure}

\usepackage{ulem}  

\def\Z2{$\mathbb{Z}_2$}
\def\I{\uppercase\expandafter{\romannumeral 1}}
\def\II{\uppercase\expandafter{\romannumeral 2}}
\def\III{{\uppercase\expandafter{\romannumeral 3}}}
\def\IV{{\uppercase\expandafter{\romannumeral 4}}}
\def\V{{\uppercase\expandafter{\romannumeral 5}}}

\def\ket#1{\vert#1\rangle}

\def\kk{\mathbf{k}}

\def\vv{\mathbf{v}}
\def\ww{\mathbf{w}}
\def\uu{\mathbf{u}}
\def\qq{\mathbf{q}}

\def\ff{\mathbf{f}}
\def\ee{\mathbf{e}}
\def\LL{\mathbf{\Lambda}}
\def\JJ{\mathbf{J}}

\def\dl{\delta\lambda}

\def\00{\mathbf{k}_0,\lambda_0}

\def\labi{LaBiTe$_3$}
\def\lubi{LuBiTe$_3$}
\def\labisb{LaBi$_{1-x}$Sb$_x$Te$_3$}
\def\lubisb{LuBi$_{1-x}$Sb$_x$Te$_3$}
\def\lasb{LaSbTe$_3$}
\def\lusb{LuSbTe$_3$}

\def\Green#1{\textcolor{OliveGreen}{#1}}

\begin{document}

\title{Weyl semimetals from noncentrosymmetric topological insulators}
\date{today}

\author{Jianpeng Liu}
\affiliation{ Department of Physics and Astronomy, Rutgers University,
 Piscataway, NJ 08854-8019, USA }

\author{David Vanderbilt}
\affiliation{ Department of Physics and Astronomy, Rutgers University,
 Piscataway, NJ 08854-8019, USA }

\date{\today}

\begin{abstract}
We study the problem of phase transitions from 3D topological to normal
insulators without inversion symmetry.
In contrast with the conclusions of some previous work,
we show that a Weyl semimetal always exists
as an intermediate phase regardless of any constriant from lattice symmetries,
although the interval of the critical
region is sensitive to the choice of path in the parameter space
and can be very narrow.
We demonstrate this behavior by carrying out
first-principles calculations on the noncentrosymmetric
topological insulators \labi\ and \lubi\ and the
trivial insulator BiTeI.
We find that a robust Weyl-semimetal phase exists in the solid
solutions \labisb\ and \lubisb\ for $x\!\approx\!38.5-41.9$\%
and $x\!\approx\!40.5-45.1$\% respectively.
A low-energy effective model is also constructed to describe
the critical behavior in these two materials. In BiTeI,
a Weyl semimetal also appears with applied
pressure, but only within a very small
pressure range, which may explain why it has
not been experimentally observed.

\end{abstract}


\maketitle


\def\scr{\scriptsize}
\ifthenelse{\equal{\draftversion}{true}}{
  \marginparwidth 2.7in
  \marginparsep 0.5in
  \newcounter{comm} 
  \def\commnext{\stepcounter{comm}}
  \def\commtext{{\bf\color{blue}[\arabic{comm}]}}
  \def\commmar{{\bf\color{blue}[\arabic{comm}]}}
  \def\dvm#1{\commnext\marginpar{\small DV\commmar: #1}\commtext}
  \def\jlm#1{\commnext\marginpar{\small JPL\commmar: #1}\commtext}
  \def\mlab#1{\marginpar{\small\bf #1}}
  \def\tnewpage{\newpage\marginpar{\small Temporary newpage}}
  \def\tfootnote#1{\Green{\scr [FOOTNOTE: #1]}}
}{
  \def\dvm#1{}
  \def\jlm#1{}
  \def\mlab#1{}
  \def\tnewpage{}
  \def\tfootnote#1{\footnote{#1}}
}

\section{Introduction}
\label{sec:intro}

The significance of topology in determining electronic properties
has became widely appreciated with the discovery of the
integer quantum Hall effect and been highlighted further by
the recent interest in
topological insulators (TIs).\cite{IQH-prl80,TKNN,kane-rmp10,zhang-rmp11}
In topological band theory,
a topological index, such as the Chern number or the \Z2\ index,
is well-defined only for gapped systems, and the topological
character is signaled by the presence of
novel gapless surface states which cannot exist
in any isolated 2D system.\cite{kane-rmp10,zhang-rmp11}
Recently, the concept of topological phases is further generalized to
3D bulk gapless systems, whose topological behavior is
protected by lattice translational
symmetry, known as the Weyl semimetal (WSM).
\cite{nielsen-plb83,burkov-prb11,wan-prb11,turner-13-review,hosur-13-review}

A Weyl semimetal is characterized by a Fermi energy that intersects
the bulk bands only at one or more pairs of band-touching points (BTPs)
between nondegenerate valence and conduction bands.  This can occur
in the presence of spin-orbit coupling (SOC), typically in a crystal
with broken time-reversal or inversion symmetry but not both, so
that the pairs are of the form ($\kk_0$, $-\kk_0$) in the Brillouin
zone (BZ).
The effective Hamiltonian around a single BTP $\kk_0$ can be written as
$H(\kk)=f_0(\kk)+\mathbf{f}(\kk)\cdot\bm{\sigma}$, where $f_0$ and $\ff$
are scalar and vector functions respectively of wavevector in the BZ
and the $\sigma_j$ are the Pauli matrices
acting in the two-band space.
If one expands the coefficient $\ff(\kk)$ to linear order around
$\kk_0$, one gets a Hamiltonian having the form of
the Weyl Hamiltonian in relativistic quantum
mechanics after a coordinate transformation in $\kk$ space.
If the sign of the determinant of the Jacobian that describes
the coordinate transformation is positive (negative), we call
the BTP as a Weyl node with positive (negative) chirality, and
the low-energy excitations around such a Weyl node provide
a condensed-matter realization of left-handed (right-handed) Weyl
fermions.

These pairs of Weyl nodes are topologically protected in the sense that
they are robust against small perturbations, which can be see from
the codimension argument as follows.
One can introduce a parameter $\lambda$
that acts as a perturbation on the BTP, and let both $f_0$ and $\ff$ to be
dependent on $\lambda$.
In order to get a band touching at $(\kk_0,\lambda_0)$,
the three coefficients $\mathbf{f}\!\equiv\!(f_x,f_y,f_z)$ have to vanish.
However, since there are four degrees of freedom,
if $\lambda_0\rightarrow\lambda_0+\dl$, instead of opening a gap,
the Weyl node would just shift slightly
in momentum to compensate for the perturbation.
In fact, there is no way to remove a Weyl node unless
two Weyl nodes with opposite chirality annihilate each other.

If the two Weyl nodes are aligned in energy
due to either time-reversal or some lattice symmetry,
and the bands are filled right
up to the Weyl nodes, then the Fermi energy
would be locked there regardless of weak perturbations. That is,
the Fermi level could be slightly shifted upward (downward) due to
some weak  perturbation, such
that there is an electron-like (hole-like) Fermi surface,
then there must also be a hole-like (electron-like) Fermi surface to
conserve the total number of electrons, which is impossible in such a
a semimetal. It follows that the low-energy physics in the Weyl semimetal
is completely dominated by the linearly dispersing states around
the Weyl nodes, which leads to interesting surface states and transport
properties.

The presence of Weyl nodes in the bulk bandstructure is responsible
for the presence of Fermi arcs at the surface, which can be
understood as follows.\cite{wan-prb11,turner-13-review} Consider
a small loop in the 2D surface BZ that encloses the projection
along $k_z$ of one Weyl point.  When translated along $k_z$,
this loop traces out a surface in the 3D BZ, and the application
of Gauss's theorem implies that the Chern number on this surface
must equal the chirality of the enclosed Weyl node.  It follows
that as $(k_x,k_y)$ is carried around the loop, a single
electron is pumped up to (or down from) the top surface, and this is
only consistent with charge conservation if a single surface
state crosses the Fermi energy $E_\textrm{F}$ during the cycle.
Since this argument applies for an
arbitrary loop, surface states must exist at $E_\textrm{F}$
along some arc emerging from the surface-projected Weyl point.
If there is another Weyl node with opposite chirality,  then the
Chern number can vanish once the cylinder encloses both of the
nodes, such that the  Fermi arc would only extend between the
two projected Weyl nodes.\cite{wan-prb11,turner-13-review}

In a WSM with broken time-reversal (TR) symmetry,
there is also a non-zero anomalous Hall
conductivity (AHC) that is closely related to the positions
of the Weyl nodes.\cite{ran-prb11,burkov-prl11,haldane-arxiv14}
Consider the AHC $\sigma_{xy}$ with the crystal oriented such that
the third primitive reciprocal vector is along $z$.
If we track the Chern number $C_z$ of a 2D slice normal to
$z$ in the 3D BZ, we must find that it changes by $\pm1$ whenever
$k_z$ passes a Weyl node, with the sign depending on the chirality
of the node.  In the simplest case,
if $C_z\!=\!1$ for $k_z$ between the two Weyl nodes and zero elsewhere,
then the AHC is just
proportional to the separation of the two Weyl nodes in $k_z$.
This is also interpreted as a consequence of
the ``chiral anomaly" in a WSM.\cite{zyuzin-prb12}
Other interesting transport phenomena can arise
due to the chiral anomaly. For example, if a magnetic field $\mathbf{B}$
is applied to a WSM in the $z$ direction, Landau levels will be
formed in the ($x,y$) plane. The zeroth Landau
level disperses linearly along $k_z$, but in opposite
directions for Weyl nodes with opposite chirality.
As a result, if an electric field $\mathbf{E}$ is
applied along $z$, electrons would be pumped from one Weyl
node to the other at a rate proportional to
$\mathbf{E}\!\cdot\!\mathbf{B}$, with the Fermi arcs serving as
a conduit.\cite{nielsen-plb83,aji-prb12,turner-13-review,haldane-arxiv14}

As discussed above, a WSM requires the breaking of either
TR or inversion symmetry.
Many of the previous works are focused on WSMs without TR symmetry,
such as in pyrochlore iridates,\cite{wan-prb11}
magnetically doped TI multilayers,\cite{burkov-prl11} and
Hg$_{1-x-y}$Cd$_x$Mn$_y$Te.\cite{bulmash-prb14}  In this paper,
we study the WSM with preserved TR symmetry
but broken inversion symmetry.

It was argued some time ago that
the \Z2-odd\ and \Z2-even\ phases of a noncentrosymmetric
insulator should always be bridged by
a critical WSM phase.\cite{murakami-prb08,murakami-njphys-07} If the
 transition is described by some
adiabatic parameter $\lambda$, then as $\lambda$ increases one
expects first the appearance of $m$ higher-order
BTPs in the half BZ (and another $m$ at the time-reversed points),
where $m\!=\!1$ is typical of low-symmetry systems while $m\!>\!1$
can occur when, e.g., rotational symmetries are present.
These higher-order BTPs generally have quadratic
dispersion in one direction while remaining linear in the other
two, and are non-chiral; we refer to such a point henceforth as
a ``quadratic BTP.''
As $\lambda$ increases, each quadratic BTP splits to form a pair
of Weyl nodes ($4m$ altogether), which then migrate
through the BZ and eventually annihilate at a second critical
value of $\lambda$ after exchanging partners.
The previous work demonstrated that this
process inverts the strong \Z2\ index if $m$ is odd.
\cite{murakami-prb08,murakami-njphys-07}  Recently,
however, Yang \textit{et al.}\ claimed that for systems with
certain high-symmetry lines in the BZ, the phase transition
could occur at a unique critical value of $\lambda$ at which
the bands would touch and immediately reopen,
instead of over some finite interval in $\lambda$,
even when inversion symmetry is absent.\cite{yang-prl13}
These authors suggested that BiTeI under pressure could serve as an
example to support
their claim.\cite{bahramy-naturecomm-12,yang-prl13}

In this paper, we address this issue carefully.
We show that an intermediate critical WSM phase should
always exist for any topological phase transition (TPT)
between a normal and a \Z2-odd insulating phase.
We find however that the width of the critical WSM phase
can be sensitive to the choice of path in parameter space
and can sometimes be very small.
To justify our conclusions, we take specific materials as examples.
We first study the TPT in the solid solutions
\labisb\ and \lubisb\ using the virtual crystal approximation,
where the phase transition is
driven by Sb substitution. The parent compounds at $x\!=\!100\%$,
\labi\ and \lubi, are hypothetical noncentrosymmetric materials
that are predicted to be strong topological insulators in
Ref.~\onlinecite{yan-prb10} and in the present work respectively.
Instead, the end members LaSbTe$_3$ and LuSbTe$_3$ at
$x\!=\!0\%$ are trivial insulators.\cite{yan-prb10}
We find that a WSM phase is obtained when $x$ is in the range of
about 38.5-41.9\% for \labi\ and 40.5-45.1\% for \lubi.
We further construct a low-energy effective model to describe
the topological and phase-transitional behavior in this
class of materials.
We also revisit the TPT of BiTeI driven by applied
pressure, where a WSM phase has not previously
been observed.\cite{bahramy-naturecomm-12,xi-prl13}
Based on our calculations, we find that
a small interval of WSM phase does actually intervene
as increasing pressure
drives the system from the trivial to the topological phase.

The paper is organized as follows. In Sec.~\ref{sec:general} we derive
the general behavior of TPTs in noncentrosymmetric insulators
and point out some deficiencies in the discussion of BiTeI by
Yang \textit{et al.}\cite{yang-prl13}
In Sec.~\ref{sec:prelim} we describe the  lattice structures
and basic topological properties of the materials, as well as the
numerical methods used in the realistic-material calculations,
especially the methods used in modeling the alloyed and
pressurized systems and in searching for BTPs in the BZ.
In Sec.~\ref{sec:materials} we present the results for \labisb,
\lubisb\ and BiTeI, and discuss the
sensitivity to the choice of path. In Sec.~\ref{sec:summary},
we summarize our work.

\section{Topological transition in noncentrosymmetric
insulators}
\label{sec:general}

\subsection{General behavior}
\label{subsec:general-nosymm}

We consider the problem of TPTs in noncentrosymmetric insulators
in the most general case. In the space of the two bands
which touch at the TPT, the system can be described by the
effective Hamiltonian
\begin{align}
H(\kk,\lambda)=f_x(\kk,\lambda)\sigma_x +
f_y(\kk,\lambda)\sigma_y
+f_z(\kk,\lambda)\sigma_z \,,
\label{equa:h}
\end{align}
where $\lambda$ is the parameter that drives the TPT and $\sigma_{x,y,z}$
are the three Pauli matrices defined in the space spanned by
the highest occupied and the lowest unoccupied states at $\kk$.
Since we study the TPT between
two insulating phases, we can assume without loss of generality that the system
is gapped for $\lambda\!<\!\lambda_0$, and that the first
touching that occurs at $\lambda\!=\!\lambda_0$ takes place at
$\kk=\kk_0$.
In other words, $f_{i}(\kk_0,\lambda_0)\!=\!0,\hspace{1pt} i\!=\!x,y,z$.
Then we ask what happens if $\kk_0\rightarrow\kk_0+\qq$ and
$\lambda_0\rightarrow\lambda_0+\dl$.

We first expand the coefficients $\ff$ around $(\kk_0,\lambda_0)$ as
$\ff\!=\!\JJ\cdot\qq+\LL\,\dl$,
where $\qq\!=\!\kk-\kk_0$, $\dl\!=\!\lambda-\lambda_0$,
$\JJ$ is the  Jacobian with matrix elements
$J_{ij}\!=\!(\partial f_i/\partial k_j)\vert_{\00}$,
and $\LL$ is a 3-vector with components
$\Lambda_i\!=\!(\partial f_i/\partial\lambda)\vert_{\00}$.
A natural set of momentum-space coordinates can be defined in
terms of the eigensystem $\JJ\cdot\vv_i=J_i\vv_i$.
Defining $\qq=\sum_i p_i\,\vv_i$ and $\uu_i=J_i\vv_i$, we obtain
\begin{equation}
\ff=\sum_{i=1}^3 p_i\,\uu_i+ \dl\,\LL .
\label{equa:forig}
\end{equation}

Following the argument
of Yang \textit{et al.},\cite{yang-prl13} the Jacobian matrix $\JJ$ has to
be singular at $(\kk_0, \lambda_0)$ because otherwise there would be
band touching even when $\lambda\!<\!\lambda_0$, contradicting the
assumption that the system is insulating for $\lambda\!<\!\lambda_0$.
This implies that at least one of the eigenvalues of $\JJ$ is zero.
We assume for the moment that the others are non-zero,
i.e., that $\JJ$ is rank-2, and let it be the first eigenvalue that
vanishes.  Since the $p_1$-dependence of $\ff$ then vanishes
at linear order, we again follow Ref.~\onlinecite{yang-prl13} by
including a second-order term to obtain
\begin{align}
\ff=p_2\,\uu_2+p_3\,\uu_3+\dl\,\LL+p_1^2\,\ww
\label{equa:f}
\end{align}
where $\ww\!=(1/2)\,\!\partial^2\ff/\partial^2p_1\vert_{\00}$.
Now we also have the freedom to carry out an arbitrary rotation
in the pseudospin representation of the two-band space.  That
is, we redefine $f_i$ to be the component in the pseudospin
in direction $\ee_i$, with $\ee_3$ given by
$(\uu_2\times\uu_3)/\vert\uu_2\times\uu_3\vert$, and
$\ee_1$ and $\ee_2$ chosen to form an orthonormal frame with $\ee_3$.
Then $u_{23}$ and $u_{33}$ vanish, and we can write explicitly that
\begin{align}
&f_1=p_2\,u_{21}+p_3\,u_{31}+\dl\,\Lambda_1 + p_1^2\,w_1,
\notag\\
&f_2=p_2\,u_{22}+p_3\,u_{32}+\dl\,\Lambda_2 + p_1^2\,w_2,
\notag\\
&f_3=\dl\,\Lambda_3+p_1^2\,w_{3} .
\label{equa:ft}
\end{align}
%

We assume $\Lambda_3/ w_3\!<\!0$, since otherwise there are solutions
at negative $\lambda$.  Then at positive $\lambda$,
 there are always two solutions
$p_1\!=\!\pm\sqrt{-\dl \Lambda_3/ w_3}$ at which $f_3\!=\!0$.
Plugging this into the expressions
for $f_1$ and $f_2$ in Eq.~(\ref{equa:ft}), we can obtain
$p_2$ and $p_3$ by solving the linear system
\begin{align}
\begin{bmatrix} u_{21} &  u_{31} \\  u_{22} & u_{32}\end{bmatrix}
\begin{bmatrix} p_2 \\ p_3\end{bmatrix}+
\begin{bmatrix} \Lambda_1 - w_1 \Lambda_3/ w_3 \\
\Lambda_2 - w_2 \Lambda_3 / w_3
\end{bmatrix}\dl=0 .
\label{equa:py-pz}
\end{align}
Solutions of the above equation always exist as long as the
Jacobian matrix
$\JJ$ is of rank two, which means a critical WSM should
always exist in the absence of a special lattice symmetry that
would lower the rank of $\JJ$.  From the above
it also follows that at the critical
$\lambda\!=\!\lambda_0$ the dispersion around $\kk_0$
is quadratic in $p_1$ and linear in $p_2$ and $p_3$, and that
for larger $\lambda$ the Weyl point displacements scale like
$\vert p_1 \vert \!\sim\!\sqrt{\dl}$ and $\vert p_{2,3} \vert\!\sim\!\dl$.
The same conclusions in the rank-two case have been
obtained by Murakami \textit{et al.}\cite{murakami-prb08} and restated by
Yang \textit{et al.}\cite{yang-prl13}

If the Jacobian matrix $\JJ$ turns out to be rank-one instead at
$\lambda_0$, then the bands would first close at a doubly-quadratic
BTP.  That is, there would be two vanishing eigenvalues of the
Jacobian matrix (which we take to be the first and second), and
the dispersion would be quadratic in $p_1$ and $p_2$ and linear
in $p_3$.  This implies that only the second-order terms associated
with $p_1$ and $p_2$ need to be included in Eq.~(\ref{equa:f}),
yielding
\begin{align}
&f_1=p_3\,u_{31}+\dl\,\Lambda_1
  + p_1^2\,w_1^{11} + p_2^2\,w_1^{22} + 2 p_1p_2\,w_1^{12} ,
\notag\\
&f_2=p_3\,u_{32}+\dl\,\Lambda_2
  + p_1^2\,w_2^{11} + p_2^2\,w_2^{22} + 2 p_1p_2\,w_2^{12} ,
\notag\\
&f_3=p_3\,u_{33}+\dl\,\Lambda_3
  + p_1^2\,w_3^{11} + p_2^2\,w_3^{22} + 2 p_1p_2\,w_3^{12} ,
\label{equa:ff}
\end{align}
where $\ww^{ij}\!=(1/2)\,\!\partial^2\ff/\partial p_i\partial p_j\vert_{\00}$
($i, j = 1, 2$).  We can make a similar transformation on $\ff$
such that the $f_3$ direction is
$\ee_3\!=\!(\uu_3\times\ww^{22})/\vert\uu_3\times\ww^{22}\vert$,
so that $f_3$ becomes independent of $p_3$ and $p_2^2$. Then one also
has the freedom to rotate the $p_1$ and $p_2$ components
to make $w^{12}_3$ vanish. After these two transformations,
$f_3$ only depends on $p_1^2$ and $\dl$, and one expects
solutions at $p_1\!=\!\pm\sqrt{-\dl\Lambda_3/w^{11}_3}$.
Plugging this into the expressions for $f_1$ and $f_2$ in
Eq.~(\ref{equa:ff}), one obtains a quadratic equation for $p_2$ of the form
$a\dl+bp_2^2+c\sqrt{\dl}p_2\!=\!0$,
where $a, b$ and $c$ are some constants determined
by the components of $\uu_3$, $\LL$, and $\ww^{ij}$ ($i,j=1,2$).
If there are real solutions
for the above equation, then the doubly-quadratic BTPs would split
into four Weyl nodes whose trajectories scale as
$p_1\!\sim\!\pm\sqrt{\dl}$ and $p_2\!\sim\!\pm\sqrt{\dl}$, $p_3\!\sim\!\dl$.
Otherwise, if there is no solution
for $p_2$, a gap would be opened up immediately after the band touching
at $(\00)$, which would represent the rare case of an
``insulator-insulator transition" using the language of
Ref.~\onlinecite{yang-prl13}.

However,  we do not expect that the
strong \Z2\ index would be inverted for such an insulator-insulator
transition in the rank-one case.  This can be seen as follows.
If the BTP does not lie in any of the TR-invariant slices
($k_j\!=\{\!0, \pi\}$, $j=1, 2, 3$),
then certainly the 2D \Z2\ indices of the TR invariant
slices would not change, and it follows that none of the four
3D \Z2\ indices would change either.
If the BTP happens to reside in one of the TR invariant slices,
then since the dispersion in the 2D slice must be
quadratic in at least one direction, it should be topologically
equivalent to the superposition
of an even number of linearly-dispersing Weyl nodes, which is
also not expected to flip the 2D \Z2\ index,
as argued in Ref.~\onlinecite{murakami-prb08}. Thus none of the
3D \Z2\ indices, including the strong index, would change.

To summarize this section, we find without any lattice-symmetry
restriction that a critical WSM phase always exists in the
rank-two case. In the rank-one case, an insulator-insulator
type transition is allowed in principle, but would not be expected
to be accompanied by a change in the strong \Z2\ index.
Therefore, it is fair to claim that, regardless of special lattice
symmetry, there is always a WSM phase connecting \Z2-odd\ and \Z2-even\
phases in a noncentrosymmetric insulator.

\subsection{Discussion of BiTeI}
\label{subsec:general-bitei}

In this section we discuss the TPT in pressured BiTeI, a case in which
the TPT is driven in a system with C$_{3v}$ symmetry. Contrary to
the conclusions of Ref.~\onlinecite{yang-prl13}, here we argue that
a critical WSM does exist in the TPT of BiTeI, although the
pressure interval over which it occurs may be rather narrow.

In Refs.~\onlinecite{bahramy-naturecomm-12,yang-prl13} the authors
argued that if there exists a high-symmetry line in the BZ such
that the dispersion extremum evolves along the line as a function
of the adiabatic parameter (pressure), then one could
get an insulator-insulator type transition without
going through a critical WSM. The authors further pointed
out that the high-symmetry lines from A to H in the BZ of
BiTeI, shown in Fig.~\ref{fig:lattice}(d), satisfy 
some necessary conditions for this to occur.
Moreover, they showed that the
symmetry of BiTeI is such that if one concentrates on the band
dispersions along these A-H lines, one finds a pair of extrema
(one valence-band maximum and one conduction-band minimum) which
migrate along the A-H line as a function of the external parameter
(pressure), coincide at a critical value, and then separate again
to reopen the gap.  They furthermore showed that the dispersions
are quadratic in the two orthogonal directions
(except exactly at the critical value), raising the
possibility that the extrema in question could be minima and
maxima in all three $k$-space directions.  This would correspond
to the insulator-insulator transition without an intervening
WSM phase.  However, our analysis in the previous section shows
that this cannot occur in the rank-two case, and that the
extrema in question actually become saddle points
after the band touchings occur along the A-H lines.
In this case, as recognized in Ref.~\onlinecite{yang-prl13},
a WSM phase does occur.
As has been verified in Ref.~\onlinecite{yang-prl13}, the Jacobian
does remain of rank two on these lines in BiTeI, and we shall show below
in Sec.~\ref{sec:bitei} that an intermediate WSM phase does occur.
We also point out that
Fig.~2 of Ref.~\onlinecite{yang-prl13} does not demonstrate the
absence of the Weyl nodes, since they are expected to lie off
the  ($k_x, k_z$) plane on which the dispersion was plotted.

Yang \textit{et al.}\cite{yang-prl13} gave another argument in
favor of the insulator-insulator scenario in BiTeI as follows.
They noted that the band touching first takes place on the A-H
line, which is invariant under the combination of time-reversal
and mirror operations. This imposes some constraints on the form
of the effective Hamiltonian around the BTP, and from these the
authors concluded that, if Weyl nodes do appear, they should
migrate along trajectories of the form $p_1\!\sim\!\pm\dl^{1/2}$,
$p_2\!\sim\!\pm\dl^{3/2}$ and $p_3\!\sim\!\dl$.  Such a curve in
3D space possesses non-zero torsion, so that the trajectories
of the two Weyl nodes emerging from one quadratic BTP could
never join again and form a closed curve.  This implies that
if the WSM is formed by such an event, then it would remain
permanently, contradicting the fact that BiTeI clearly becomes a
globally-gapped TI at higher pressures.  Based on this reasoning,
they concluded that the TPT in BiTeI must be an insulator-insulator
transition without an intermediate WSM.

However, this argument neglects the fact that the C$_{3v}$
symmetry means that there are several A-H lines in the BZ of
BiTeI, and the gap first closes by the simultaneous appearance of
quadratic BTPs at equivalent positions on all of these lines.
Even though the two Weyl nodes which emerge from a single
quadratic BTP cannot meet each other, as shown from the torsion
of their trajectories, the Weyl nodes from different BTPs can
interchange partners and eventually annihilate each other in
such a way as to form a closed curve in the BZ.  This is exactly
the mechanism of the topological phase transition in noncentrosymmetric 
TIs.\cite{murakami-njphys-07,murakami-prb08} As will
be discussed in Sec.~\ref{sec:bitei}, there are actually six
quadratic BTPs in the full BZ that appear simultaneously,
according to the crystalline and TR symmetries.  These six Dirac
nodes split into twelve Weyl nodes, which are eventually gapped
out by annihilation after exchanging partners.

In the following section, we will study the TPTs in various inversion
asymmetric materials by first-principles calculations.
We  predict \labisb\ and \lubisb\ to be WSM
candidates within a certain range of impurity composition $x$.
We also revisit the case of BiTeI, and find that a
WSM phase emerges when external pressure
is applied to BiTeI, but only within a small pressure interval.

\section{Preliminaries}
\label{sec:prelim}

\subsection{Lattice structures and basic topological properties}
\label{sec:prelim-basic}

\begin{figure}
\includegraphics[width=7.5cm]{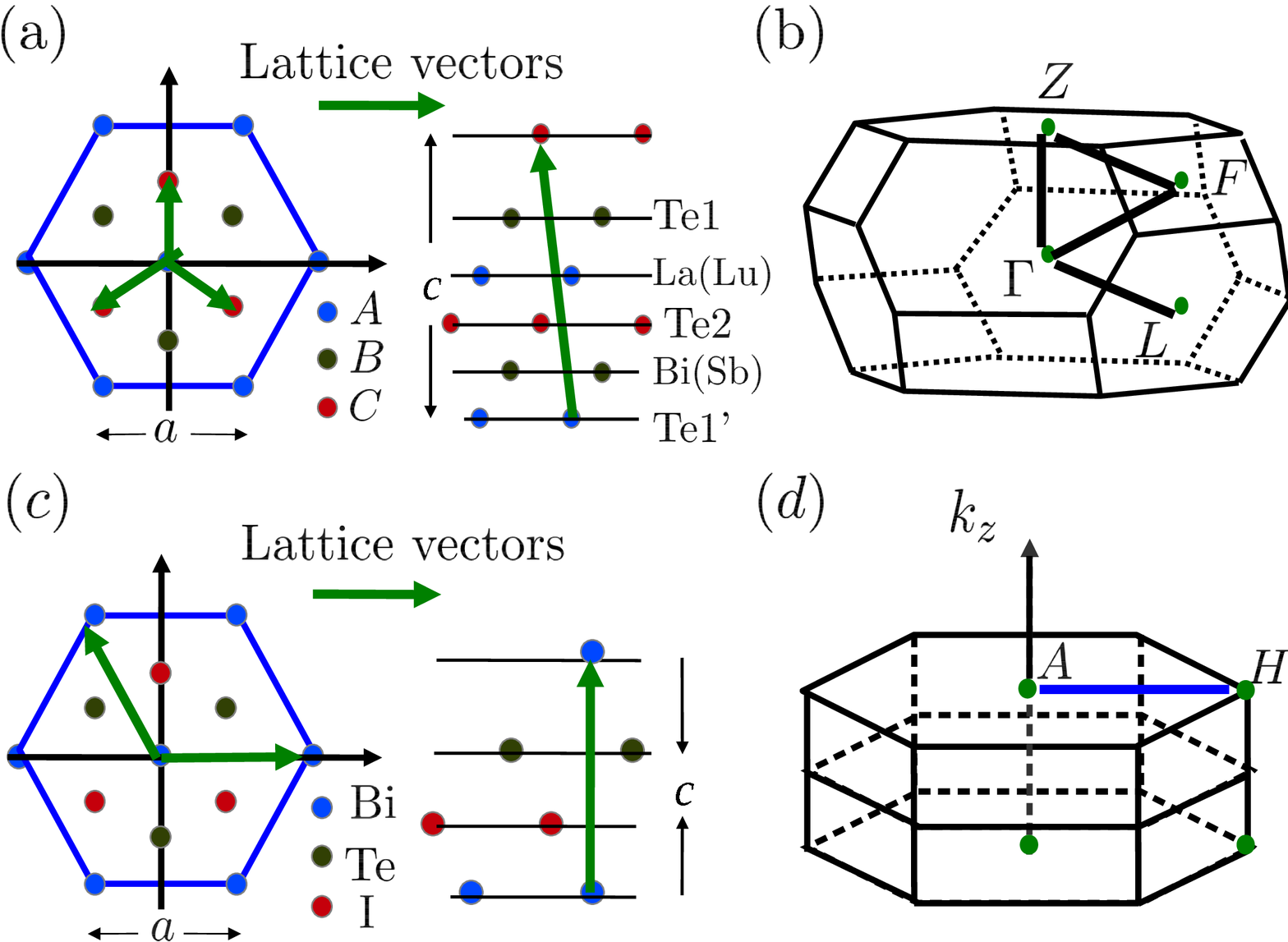}
\caption{(a) The lattice structure of \labi, \lubi, \lasb, and \lusb.
(b) The BZ of La(Lu)Bi(Sb)Te$_3$.
(c) The lattice structure of BiTeI. (d) The BZ of BiTeI.}
\label{fig:lattice}
\end{figure}


The assumed crystal structures of \labi\ and \lubi\ are very similar to
Bi$_2$Te$_3$, where five atomic monolayers stack in the
[111] direction in an \hbox{$A$-$B$-$C$-$A$-...}
sequence forming quintuple layers (QLs) as shown in
Fig.~\ref{fig:lattice}(a).  The only difference is that one of
the two Bi atoms in the primitive unit cell is replaced by a La or Lu atom,
which breaks the inversion symmetry.
The lattice structure of \lasb\ and \lusb\ is the same
as for \labi\ and \lubi, except that all the Bi atoms are substituted by
Sb. The in-plane hexagonal lattice parameters
for \labi\ and \lubi\ are $a\!=\!4.39$\,\AA\ and 4.18\,\AA\ respectively,
while the size of a QL along
$c$ is 10.07\,\AA\ and 10.29\,\AA\ respectively.
The lattice parameters of \lasb\ are slightly
different from \labi, with $a\!=\!4.24$\,\AA\ and $c\!=\!10.13$\,\AA.
The lattice parameters for \lusb\ have not been reported before,
so we use those from \lubi.
Among these four hypothetical materials, \labi\ has been
previously reported as a candidate for an inversion-asymmetric
TI.\cite{yan-prb10} \lubi\ is first reported as
a TI candidate in this paper; the non-trivial band topology
is confirmed by calculating the bulk \Z2\ index\cite{soluyanov-prb11}
and checking the existence of topological surface states.
On the other hand, \lasb\ and \lusb\ are trivial insulators.

As shown in Fig.~\ref{fig:lattice}(c),
BiTeI has a hexagonal lattice structure with three atoms
in the primitive cell stacked as \hbox{$A$-$B$-$C$-$A$-...}
along the $z$ direction. The lattice parameters in-plane and
along the hexagonal axis are $a\!=\!4.339$\,\AA\ and
$c\!=\!6.854$\,\AA. BiTeI itself is a trivial insulator
with a large Rashba spin splitting in the bulk,\cite{bitei-rashba}
but it can be driven into a TI state by applying pressure.
Previous studies have suggested that the transition to the
topological phase is not mediated by a WSM phase,
\cite{bahramy-naturecomm-12,yang-prl13}
but we revisit this issue in Sec.~\ref{sec:bitei} and
come to different conclusions.

\subsection{First-principles methodology}
\label{sec:prelim-method}

We carry out the bulk first-principles calculations
using the VASP package including SOC.\cite{vasp1,vasp2}
The generalized-gradient approximation
is used to treat the exchange-correlation functional.\cite{pbe-1,pbe-2}
The BZ is sampled on an
8$\times$8$\times$8 Monkhorst-Pack\cite{monkhorst-pack}  $\kk$ mesh
and an energy cutoff of 340\,eV is used.
The output from the first-principles plane-wave calculations
are then interfaced to the Wannier90 package\cite{wannier90}
to construct realistic tight-binding (TB) models for
these materials.\footnote{
  The TB models from Wannier90 are realistic in the sense
  that the Wannier-interpolated bandstructure can reproduce
  the first-principles band energies exactly in a specified
  energy window which, is chosen here to be centered around
  the Fermi level.}

To describe the electronic structure of \labisb\ and \lubisb,
we adopt the virtual crystal approximation (VCA) in which
each Bi or Sb is replaced by a ``virtual" atom
whose properties are a weighted average of
the two constituents. The VCA treatment typically
gives a reasonable description for solid-solution systems
in which the dopant and host atoms have a similar
chemical character.  For example, the VCA was shown to work well
in describing Sb substitution in Bi$_2$Se$_3$, because of the similar
orbital character of Sb $5p$ and Bi $6p$, but not for In
substitution, where In $5s$ orbitals become involved.\cite{jpl-prb13}
The VCA is implemented in the Wannier basis
by constructing separate 36-band models for \labi\ (\lubi)
and \lasb\ (\lusb), including all the valence
$p$ orbitals of the cations and anions, as well as
the $5d$ and $6s$ orbitals of the rare-earth elements.\footnote{The
  fully occupied $f$ shell of Lu has very little influence on
  the electronic structure around the Fermi level.}
In the solid solution, the Hamiltonian matrix elements
are then taken as a linear interpolation
in impurity composition $x$ of the corresponding matrix elements
of the parent materials.  That is, we take
$H_{mn}^\textrm{VCA}=(1-x)H_{mn}^\textrm{Bi}+xH_{mn}^\textrm{Sb}$,
where $H_{mn}^\textrm{Bi}$ and $H_{mn}^\textrm{Sb}$
denote the matrix elements of the TB models of \labi\ and \lasb.
It worth noting that when generating the WFs
for the VCA treatment,  the Wannier basis
functions have to be chosen as similar as possible before the
averaging.\cite{jpl-prb13} We therefore use
WFs that are constructed simply by projecting the
Bloch states onto the same set
of atomic-like trial orbitals without applying a subsequent
maximal-localization procedure.\cite{MLWF-1, MLWF-2}

Similarly, to study the pressure-induced TPT in BiTeI, we carry out
first-principles calculations for the system at the zero-pressure
volume, where it is topologically normal, and also at 85.4\% of
the original volume,
a value chosen somewhat arbitrarily to be well inside the TI region.
\cite{bahramy-naturecomm-12}
We denote these two states as $\eta\!=\!0$ and $\eta\!=\!1$ respectively.
Then from the
Wannier representation we again construct a realistic Hamiltonian
for each system, denoted as $H_0$ and $H_1$ respectively,
including all the valence $p$ orbitals of Bi, Te and I.  Finally we
linearly interpolate these as $H(\eta)\!=\!(1-\eta)H_0+\eta H_1$,
treating $\eta$ as an adiabatic parameter that tunes the system
through the topological phase transition.

Using these Wannierized effective TB models, we can search for
BTPs very efficiently over the entire BZ. We first sample the
irreducible BZ using a relatively sparse $\kk$ mesh, e.g.,
20$\times$20$\times$20, and find the point $\kk_0$ having the
smallest direct band gap on this mesh.  A second-round search is
conducted by scanning over a denser $\kk$ mesh within a sphere
centered on $\kk_0$.  We then repeat the procedure iteratively
until convergence is reached.  All of the trajectories of Weyl
nodes presented in Sec.~\ref{sec:materials} are obtained using
this approach.

\section{Results}
\label{sec:materials}

\subsection{\labisb\ and \lubisb}
\label{sec:labisb}

%
\begin{figure}
\centering
\includegraphics[width=6.5cm]{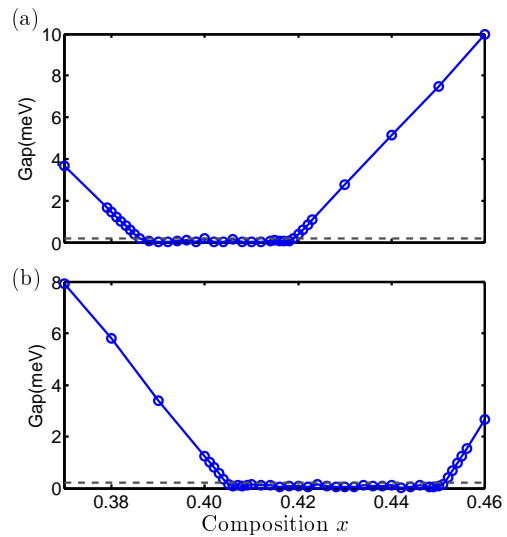}
\caption{Smallest direct band gap in the BZ vs.\ composition $x$
for (a) \labisb\ and (b) \lubisb. Dashed gray line marks
our chosen threshold of 0.2\,meV to signal a gap closure. }
\label{fig:gap}
\end{figure}

\subsubsection{Band gap and Weyl chirality}

\def\lao{38.5}
\def\lat{41.9}
\def\luo{40.5}
\def\lut{45.1}

For each of these materials we scan over a mesh in composition
$x$, and for each $x$ we construct the Wannierized Hamiltonian
for the corresponding solid solution within the VCA.
We then use the methods of the previous section to search for the
BTPs in the entire irreducible BZ.  Plots of the smallest direct
band gap in the BZ vs.\ $x$ are presented in Fig.~\ref{fig:gap}.
Clearly the gap remains closed over a finite range of $x$ in both
cases, from \lao\% to \lat\% for \labisb\  and
\luo\% to \lut\% for \lubisb.  By checking the dispersion
around the gap-closure point, we confirm that the system is
semimetallic with the Fermi level lying at a set of degenerate
Weyl BTPs over this entire range.

\begin{figure}
\centering
\includegraphics[width=6.5cm]{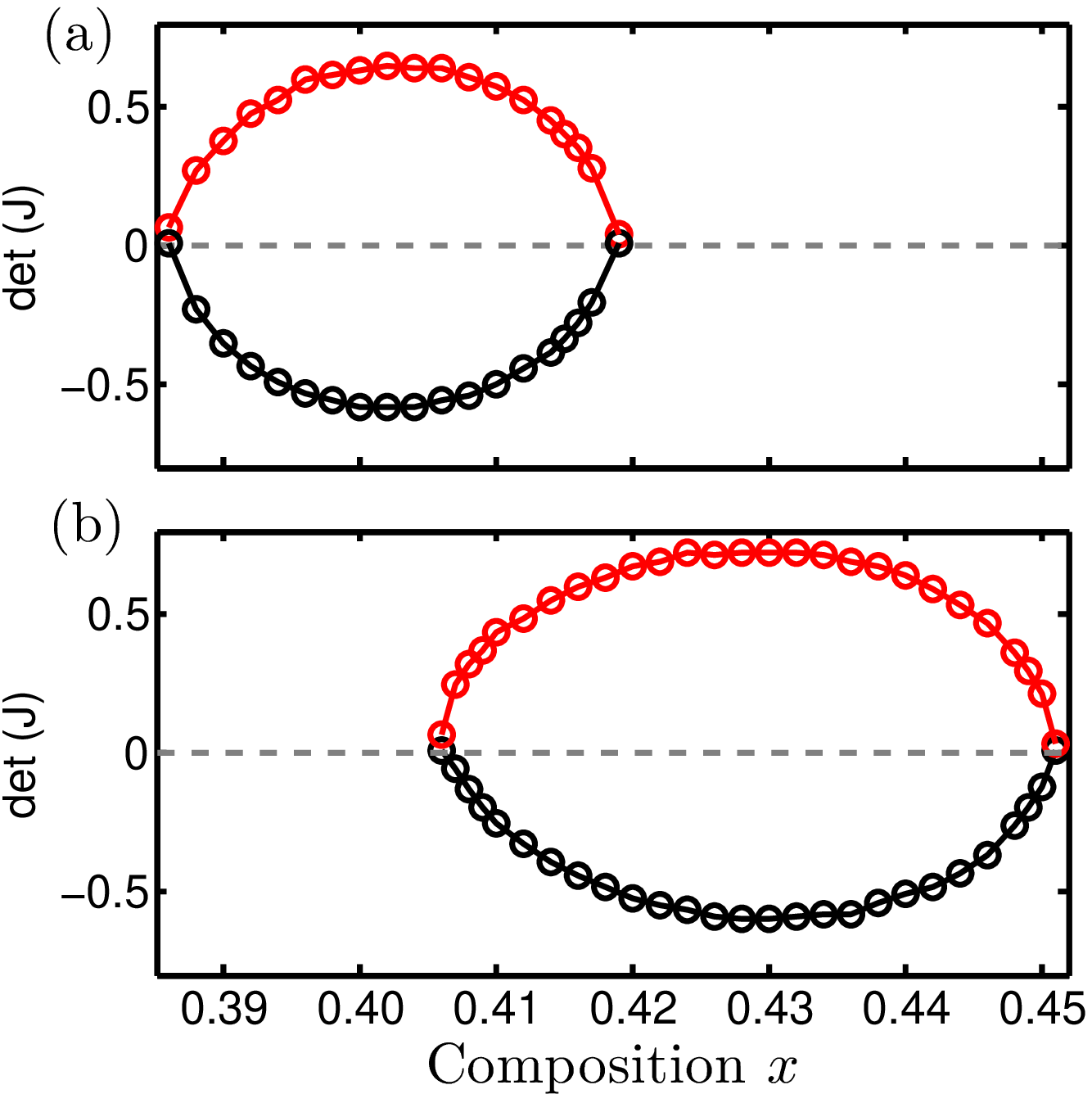}
\caption{ Determinant of the Jacobian matrix evaluated at Weyl
nodes with positive (red) and negative (black) chirality
vs.\ composition $x$, for (a) \labisb\ and (b) \lubisb.}
\label{fig:chiral}
\end{figure}

To illustrate the topological character, we further calculate the
chirality of the BTPs, which
is given by the determinant of the Jacobian matrix
$J_{ij}=\partial f_i/\partial k_j$. Fig.~\ref{fig:chiral}
shows how $\det(\JJ)$ varies with $x$ for the BTPs in
\labisb\ and \lubisb. The red  and black open circles mark the values
of $\det(\JJ)$ for the BTPs with positive and negative chirality,
which are mapped into each other by mirror operations about the $k_x\!=\!0$
and other equivalent mirror planes.  One can see that at the
beginning of the band touching, the chirality starts at zero,
indicating the creation of a quadratic BTP.  As $x$ increases, each
quadratic BTP splits into two Weyl nodes with opposite chirality.
These then migrate through the BZ and eventually annihilate each
other at the point where the chirality returns to zero.

\begin{figure}
\centering
\subfigure{\includegraphics[height=5cm]{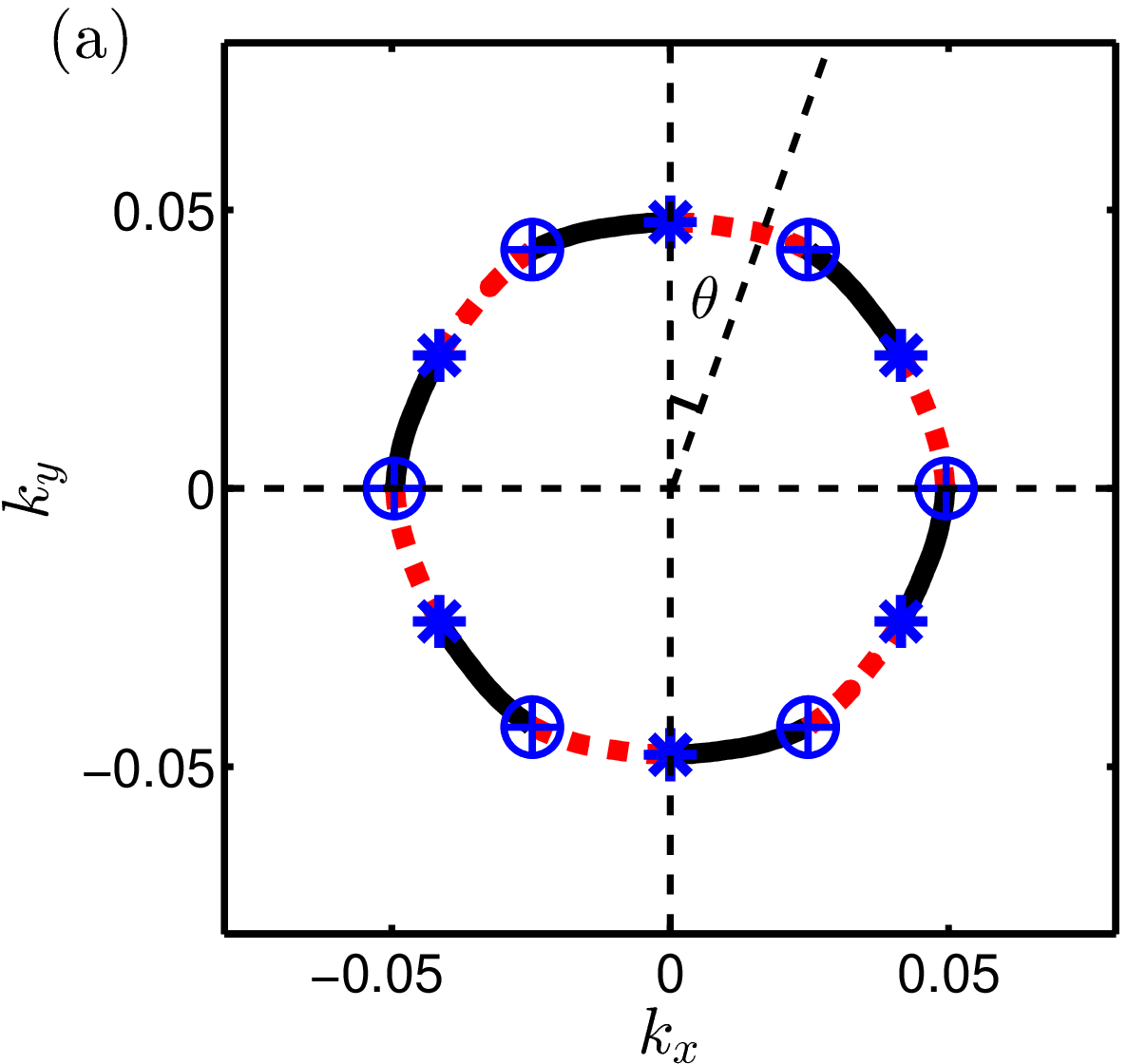}}
\subfigure{\includegraphics[height=5cm]{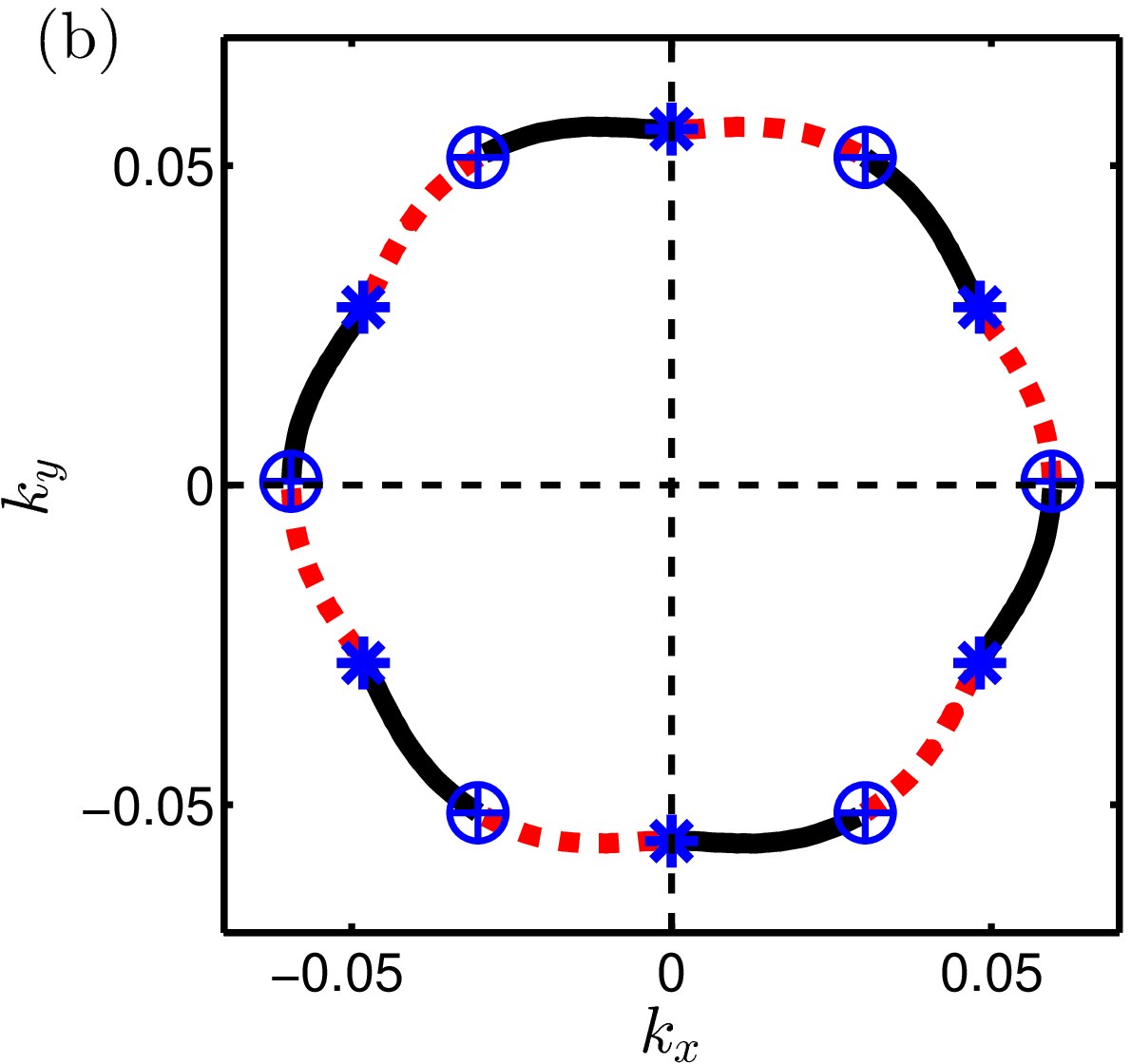}}
\caption{Trajectories of Weyl nodes in the ($k_x, k_y$) plane
(in units of \AA$^{-1}$). Dashed red lines indicated Weyl nodes
of positive chirality; solid black lines are negative.
The `*' and `$\oplus$' denote respectively the points of
creation or annihilation of Weyl nodes.
(a) For \labisb.
(b) For \lubisb.}
\label{fig:kxky}
\end{figure}

\subsubsection{Symmetry considerations}

As mentioned earlier,
the point group of this class of materials is C$_{3v}$,
which has a 3-fold rotation axis along $k_z$ and three mirror
planes that contain the $k_z$ axis and
intersect the $k_z\!=\!0$ plane on the lines $k_x\!=\!0$
and $k_y\!=\!\pm k_x/\sqrt{3}$.  We define an azimuthal angle
$\theta$ that measures the rotation of $(k_x,k_y)$ from the
$+k_y$ axis in the clockwise direction as shown in
Fig.~\ref{fig:kxky}(a).
As a result of the three-fold rotational symmetry, if a Weyl node
with positive chirality appears at some $\theta$ in the
region $0\!\leq\theta\!\leq\pi/3$ and at some $k_z$ in the
upper half BZ, then there must be another two nodes with the same
chirality and the same $k_z$ located at
$\theta+2\pi/3$ and $\theta-2\pi/3$.
Taking into account the mirror symmetry, these must have
negative-chirality partners at the same $k_z$ but at
$-\theta$, $-\theta+2\pi/3$ and $-\theta-2\pi/3$.
Finally, because of TR symmetry, each
Weyl node at $\kk$ is always accompanied by another at $-\kk$
with the same chirality, giving six more Weyl nodes
in the lower half BZ. We thus generically
expect a total of twelve Weyl nodes in the entire BZ for
compositions $x$ in the region of the WSM phase.

\subsubsection{Weyl trajectories}

Figure ~\ref{fig:kxky} shows the trajectories of the Weyl
nodes in \labisb\ and \lubisb\ projected onto the ($k_x, k_y$) plane
as $x$ passes through the critical region.
The red dashed line represents the trajectory of Weyl nodes with
positive chirality, while the solid black one denotes those with
negative chirality, and the ``*" and ``$\oplus$" denote the creation and
annihilation points of the Weyl nodes respectively.
As $x$ increases, six quadratic BTPs  are simultaneously created
in the mirror planes; this occurs at $x_{\rm c1}\!=\!\lao\%$ for
\labisb\ and $\luo\%$ for \lubisb.
Each quadratic
BTPs then splits into two Weyl nodes of opposite chirality, and
these twelve nodes migrate along the solid black and dashed red
lines shown in the figure.
Eventually, after exchanging partners, the Weyl nodes meet
and annihilate each other in another set of high-symmetry planes
($k_y\!=\!0$ and other equivalent planes), at $x_{\rm c2}\!=\!\lat\%$
for \labisb\ and $\lut\%$ for \lubisb.

\begin{figure}
\centering
\includegraphics[width=8.6cm]{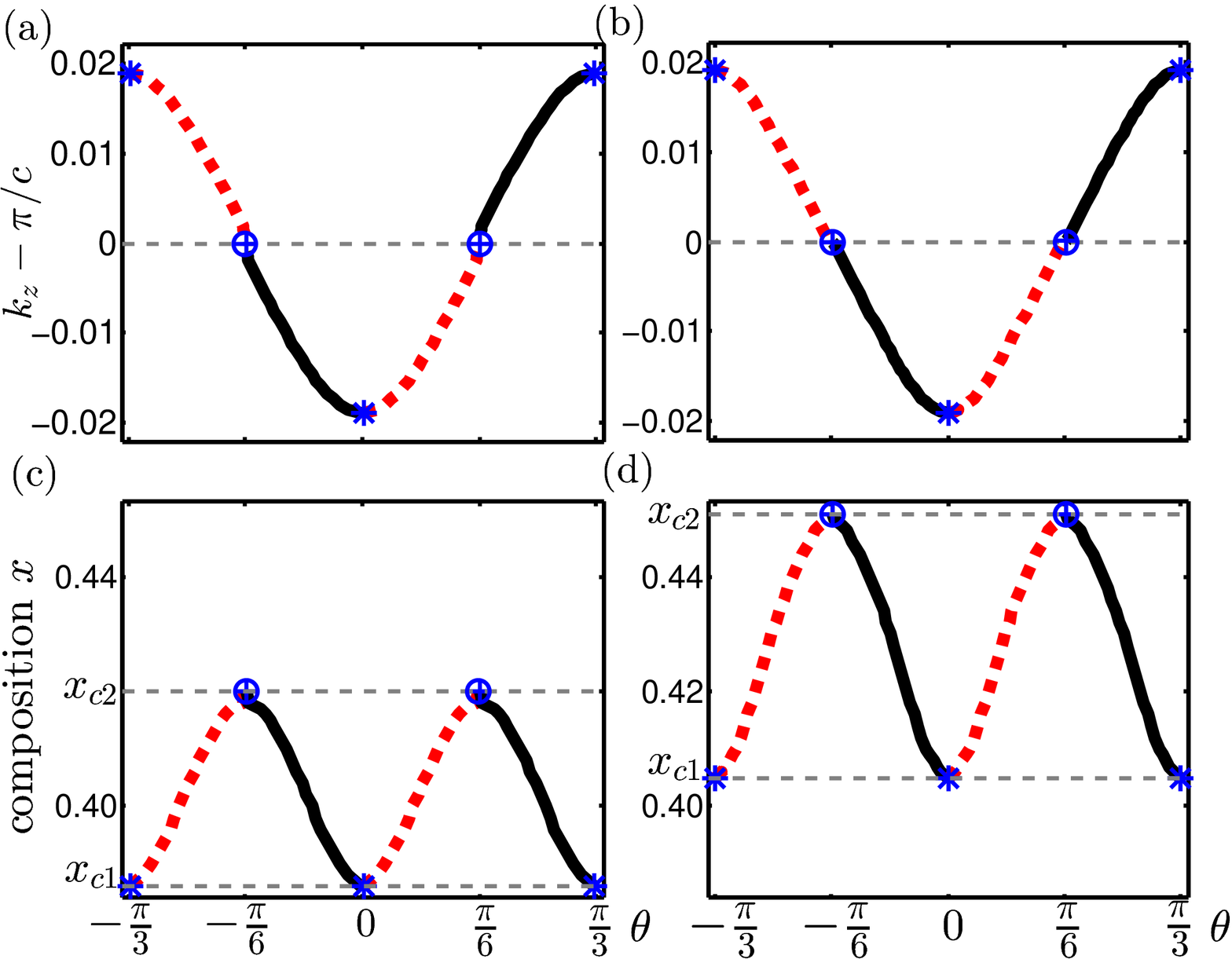}
\caption{(a-b): Trajectories of Weyl nodes in the $k_z$ direction
(in units of \AA$^{-1}$) for (a) \labisb\ and (b) \lubisb.
Dashed red (solid black)
lines refer to the Weyl nodes with positive (negative) chirality. 
$\theta$ is the azimuthal angle in the ($k_x, k_y$)
plane, as indicated in Fig.~\ref{fig:kxky}(a).
The ``*" and ``$\oplus$" denote the creation and annihilation point
of the Weyl nodes respectively.
(c-d): Trajectories of Weyl nodes in the direction of impurity 
composition $x$ for (c) \labisb\ and (d) \lubisb.}
\label{fig:xkz}
\end{figure}

Figure~\ref{fig:xkz}(a)-(b) shows the trajectory of the Weyl nodes
in the $k_z$ direction. At $x\!=\!x_{\rm c1}$, six quadratic BTPs
are created, three in the top half-BZ and three in the bottom half-BZ,
but all of them fairly close to the BZ boundary plane at
$k_z\!=\!\pm\pi/c$.  As $x$ increases, the six
BTPs split to form twelve Weyl nodes, and these begin to move
toward the above-mentioned BZ boundary plane.
Finally, after interchanging partners, Weyl nodes of opposite
chirality annihilate in pairs at $x_{\rm c2}$ on the
BZ boundary plane at $k_z\!=\!\pm\pi/c$.  For $x>x_{\rm c2}$
a global gap opens up and the system is again an insulator but
with an inverted \Z2\ index.

The locus of Weyl points can be regarded as forming a loop in the
4D space of ($k_x,k_y,k_z,x$), and just as this loop can be projected
onto $k_z$ as in Figs.~\ref{fig:xkz}(a-b), it can also be
projected onto the direction of impurity composition
$x$ as shown in Figs.~\ref{fig:xkz}(c-d). Again, it is clear
that the Weyl nodes are created at $x_{c1}$ in the mirror planes
and annihilated at $x_{c2}$ at $\theta\!=\!\pm\pi/6$. These plots may
also be helpful in seeing how the high six-fold symmetry
contributes to the narrowness of the WSM region.
If the symmetry of the system were lower,
the period of oscillation in $\theta$ in Figs.~\ref{fig:xkz}(c-d) would
be longer, which would allow the Weyl nodes to oscillate 
farther in the $x$ direction, giving a wider window of 
concentration for the WSM phase. In contrast, a fictitious system 
with an $N$-fold rotational symmetry would force the width of the 
WSM region to vanish as $N\rightarrow\infty$.  Here we have $N\!=\!6$, 
which is evidently large enough to limit the WSM phase to a rather 
small interval in $x$.

%

\subsubsection{Surface Fermi arcs}

\begin{figure}
\centering
\subfigure{\includegraphics[width=6cm]{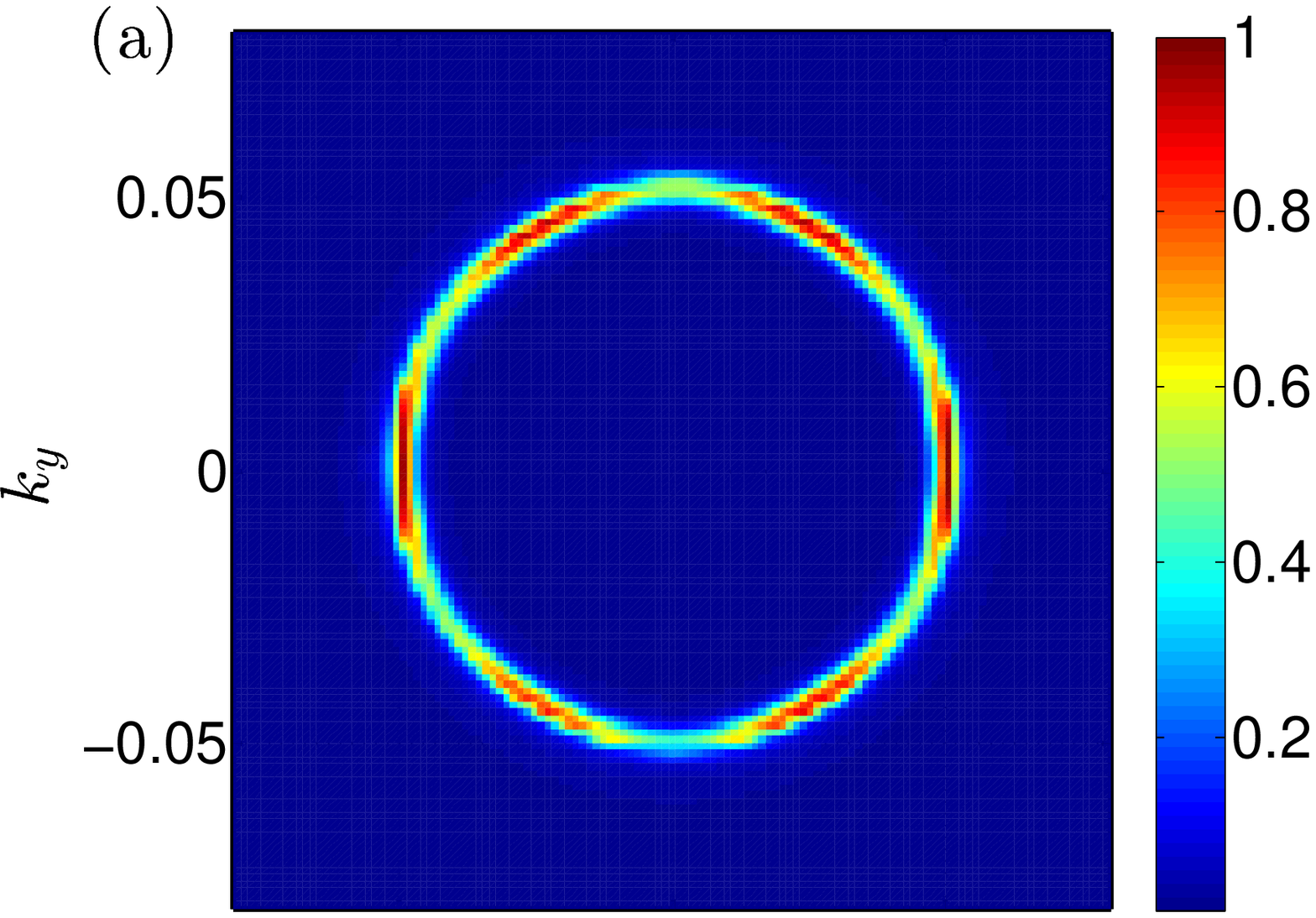}}
\subfigure{\includegraphics[width=6cm]{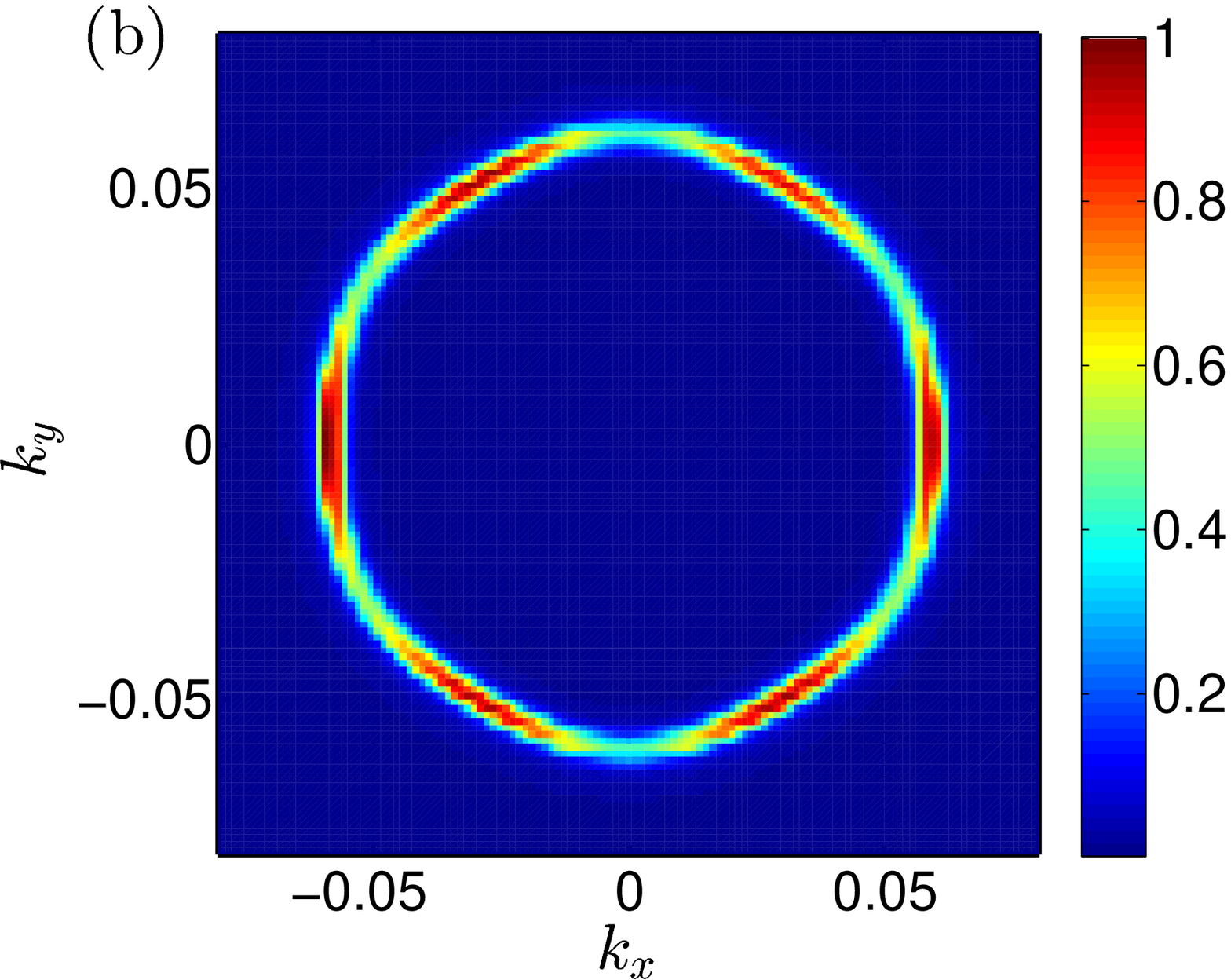}}
\caption{Surface spectral function averaged around the Fermi level
($k_x$ and $k_y$ in units of \AA$^{-1}$) for
(a) \labisb\ at $x\!=\!0.405$,
(b) \lubisb\ at $x\!=\!0.43$.}
\label{fig:surface}
\end{figure}

One of the most characteristic features of WSMs is the existence of
Fermi arcs in the surface bandstructure.
Here we calculate the surface states using the surface
Green's-function technique,\cite{surface-gf}
which is implemented in the context of the VCA effective Hamiltonian
in the Wannier basis. The surface BZ is sampled by a
64$\times$64 $\kk$ mesh, and the surface spectral functions calculated
on this mesh are then linearly interpolated to fit a
128$\times$128 $\kk$ mesh. Fig.~\ref{fig:surface} shows the normalized 
surface spectral functions averaged around the Fermi level for \labisb\ 
at $x\!=\!0.405$ and for \lubisb\ at $x\!=\!0.43$.
The averaging is done over an energy window of $\pm4.5\,$meV
around the Fermi energy, which
is determined by the position of the bulk Weyl nodes.
Six Fermi arcs connecting the projected
Weyl nodes of opposite chirality are visible, confirming the existence
of the WSM phase in these two solid-solution systems.
Note that because of the small projected bulk gap
on the loops where the Fermi arcs reside, some non-negligible
spectral weight is visible even outside the Fermi arcs in
Fig.~\ref{fig:surface}, coming from the artificial smearing of
the Green's functions.
\begin{figure}
\centering
\subfigure{
\includegraphics[width=6.0cm]{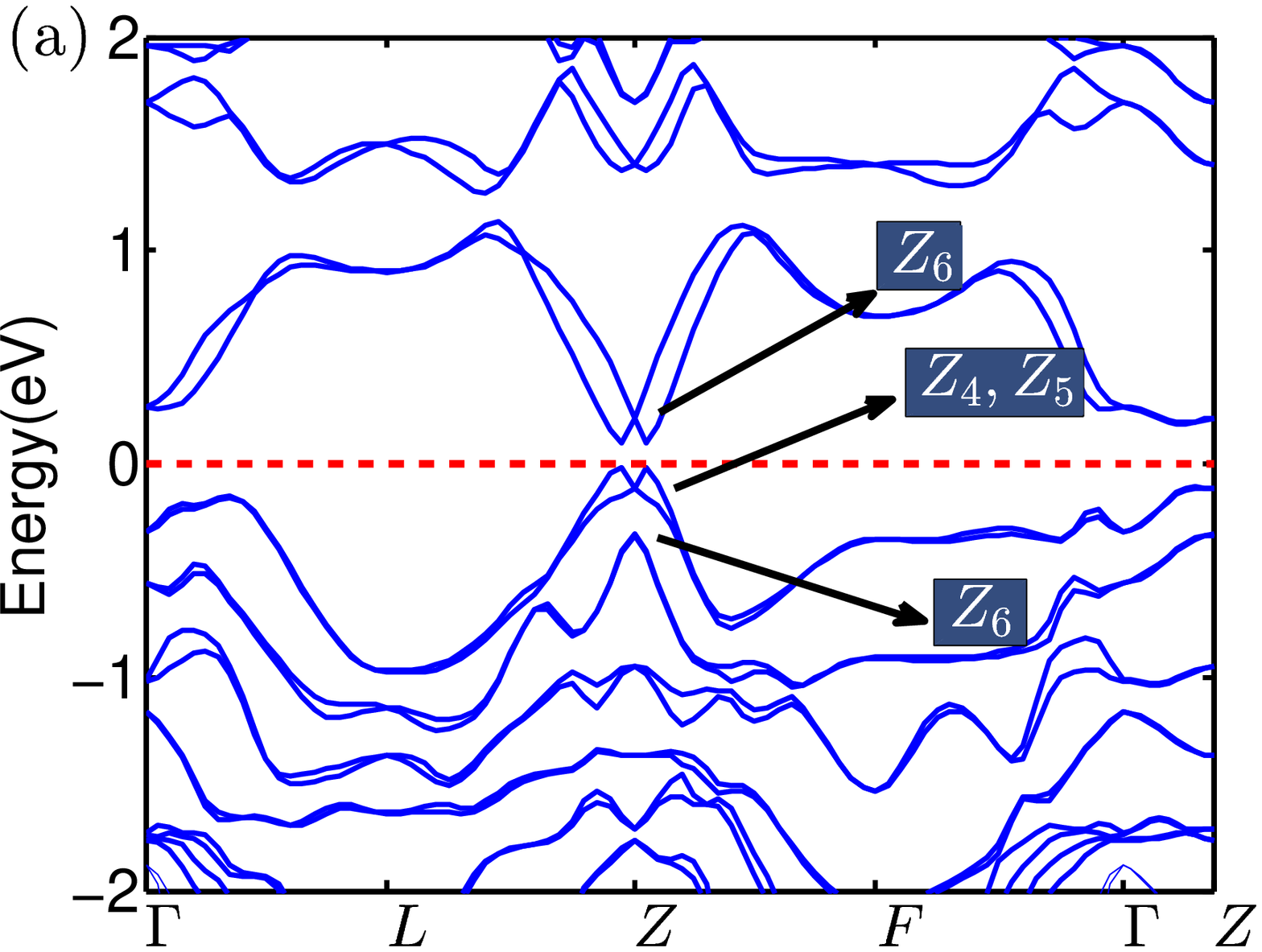}}
\vspace{0.5cm}
\subfigure{
\includegraphics[width=6.0cm]{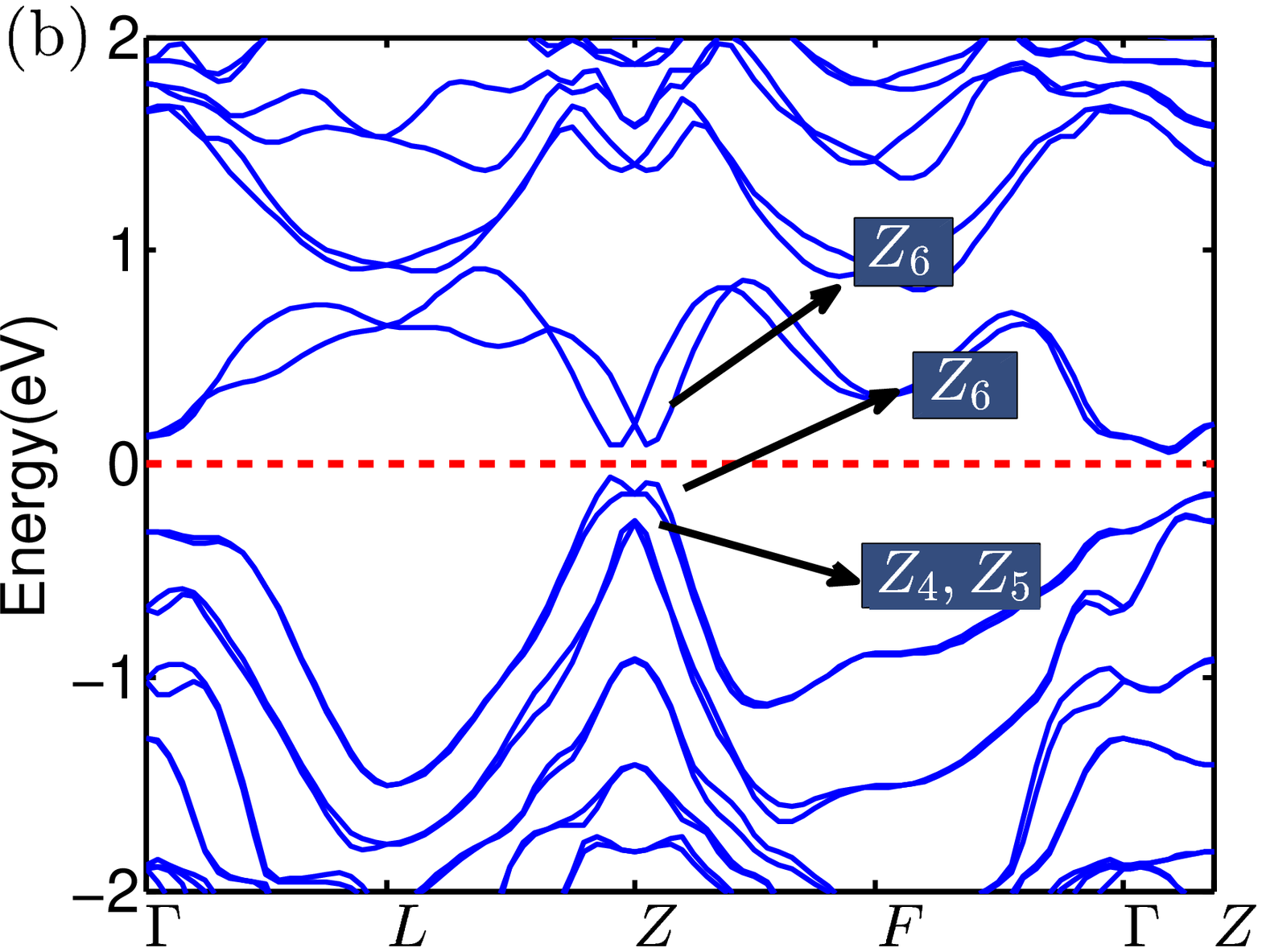}}
\caption{Bulk bandstructures of (a) \labi\ and (b) \lubi.}
\label{fig:bulk-band}
\end{figure}
%

\subsubsection{Simplified six-band model}

In order to capture the essential physics in these materials, we
construct a six-band TB model to describe the interesting
critical behavior.  From the bandstructures plots presented in
Fig.~\ref{fig:bulk-band}, it is clear that the band inversion occurs
around the $Z$ point of \labi\ and \lubi,
so we focus our attention on the six states at
$Z$ closest to the Fermi level.
A symmetry analysis shows that these six states belong to two
copies of the two-dimensional $Z_6$ irreducible representation (irrep)
of the C$_{3v}$ group at $Z$, plus a Kramers pair of 
one-dimensional complex-conjugate $Z_4$ and $Z_5$ irreps 
corresponding to linear combinations of $j_z\!=\!\pm 3/2$ orbitals.

\begin{figure}
\includegraphics[width=7cm,clip]{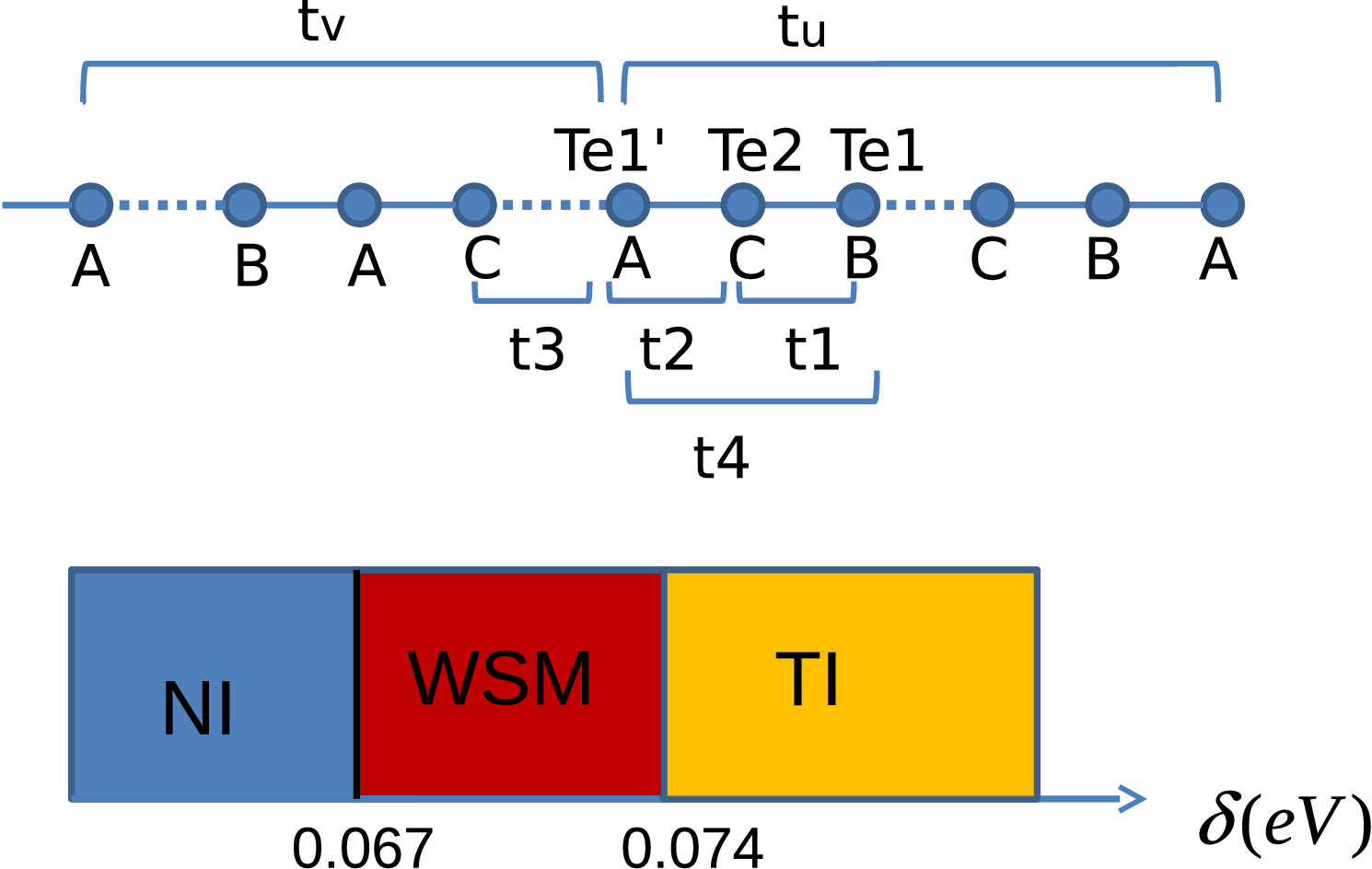}
\caption{Top: Schematic diagram of the inter-layer
spin-independent hopping terms
in the six-band model. Orbitals on sites Te1, Te2, and Te1$'$
make up a quintuple layer; A, B and C label in-plane hexagonal positions.
Bottom: Phase diagram for the topological behavior of the six-band model.}
\label{fig:6band1}
\end{figure}

We thus build our six-band TB model out of basis states having the
symmetry of $\ket {p_z,\uparrow}$ and $\ket {p_z,\downarrow}$ on
the Te atoms at the top and bottom of the quintuple layer, and
$\ket {p_x+ip_y,\uparrow}$ and $\ket {p_x-ip_y,\downarrow}$
combinations located on the central Te atoms.
A schematic illustration of the
six-band model is shown in Fig.~\ref{fig:6band1},
where the top, bottom and central Te atoms are denoted by Te$1$, Te$1'$
and Te$2$ respectively.
First of all, six inter-layer spin-independent
hopping terms are included in the model.
As shown in Fig.~\ref{fig:6band1}, we consider the first-neighbor
hopping between the central and top (bottom) Te atoms $t_1$ ($t_2$),
the inter-QL (intra-QL) hopping between the top and bottom
Te atoms $t_3$ ($t_4$), and some further-neighbor
hoppings $t_u$ and $t_v$ that are crucial in obtaining a
nontrivial \Z2\ index.
Second, to capture the Rashba spin-splitting in the
first-principles bandstructure, in-plane Rashba-like
spin-dependent hoppings within the top and bottom Te monolayers
are included and are denoted by $\lambda_{1}$ and
$\lambda_{2}$ respectively.  For completeness, the inter-layer
first-neighbor Te$1$-Te$2$ ($\lambda_{3}$) and Te$1'$-Te$2$ ($\lambda_{4}$)
Rashba-like hopping terms are also included.
Lastly, to reproduce the first-principles bandstructure better,
we also introduce first-neighbor spin-independent
hopping terms within the Te1, Te2 and Te$1'$ monolayers,
denoted by $v_1$, $v_2$ and $v_3$ respectively.
The onsite energies are also different and are labeled by
$E_1$ for Te1, $E_2$ for Te2, and $E_3$ for Te$1'$.
As our model is only intended to be
semiquantitative, we use the same model parameters to describe
both  \labi\ and \lubi.

We take all of the parameters in the model to depend a scaling
parameter $\delta$ that drives the TPT. When $\delta$ is zero,
the system is a trivial insulator; as $\delta$ increases,
the system becomes a topological insulator by going through
a critical WSM.
The dependence of the parameters on $\delta$ defines a path
in parameter space. It is important to note that
the width of the critical WSM region can be highly sensitive
to this path, with an improper choice sometimes leading to an
extremely narrow WSM phase.
Our choice is specified in Table~\ref{table:parameter}.

\begin{table}[b]
\caption{Parameters of the six-band model (in eV).}
\begin{ruledtabular}
\begin{tabular}{lclclc}
$t_1$ & $0.2-\delta/4$ &
      $\lambda_3$ & $0.15-\delta/2$ &
           $v_3$  & 0 \\
$t_2$ & $0.15-\delta/4$ &
      $\lambda_4$ & $0.12-\delta/2$ &
           $E_1$ & $0.1+\delta-6v_1$ \\
$t_3$ & $\delta$ &
      $t_u$ & $0.12+\delta/2$ &
           $E_2$  & $-6v_2$ \\
$t_4$ & $0.1-\delta/4$  &
      $t_v$ & $0.06-\delta/2$ &
           $E_3$  & $-0.1-\delta$ \\
$\lambda_1$ & $0.24-\delta/2$ &
      $v_1$ & 0.05 & & \\
$\lambda_2$ & $0.2-\delta/2$ &
      $v_2$ & 0.1 & & \\
\end{tabular}
\end{ruledtabular}
\label{table:parameter}
\end{table}

\begin{figure}
\includegraphics[width=8.5cm,clip]{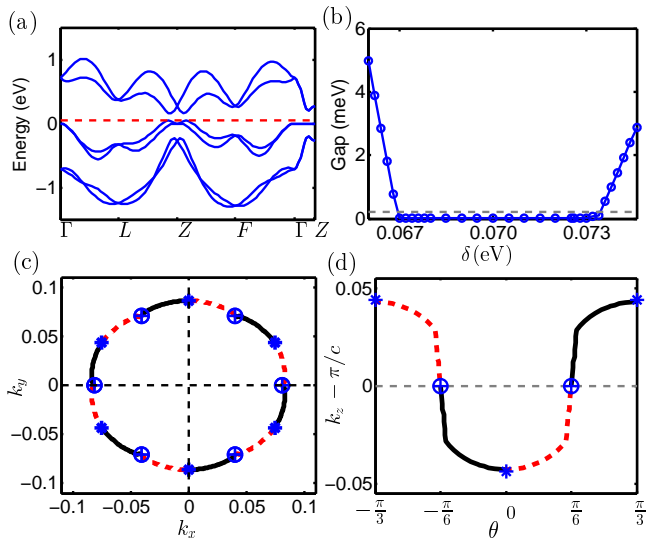}
\caption{(a) Bulk bandstructure of the six-band model at
$\delta\!=\!0.09$\,eV. (b) Smallest direct band gap in the BZ
vs.~$\delta$. (c) Trajectory of Weyl nodes projected onto the
($k_x, k_y$) plane.  Dashed red (solid black)
line refers to the Weyl node with positive (negative) chirality.
The ``*" and ``$\oplus$" denote the creation and annihilation point
of the Weyl nodes respectively.
$\theta$ is the azimuthal angle in the ($k_x, k_y$) plane.
(d) Trajectory of Weyl nodes along $k_z$.
Units of $k_x$, $k_y$ and $k_z$ are \AA$^{-1}$.}
\label{fig:6band-all}
\end{figure}

Following the path we have chosen, a WSM phase is obtained
for $0.067$\,eV\,$<\!\delta\!<\!0.074$\,eV.
As shown in Fig.~\ref{fig:6band-all}(b), the smallest
direct band gap in the BZ vanishes when
$0.067$\,eV$\,<\!\delta\!<\!0.074$\,eV, indicating the existence
of BTPs in BZ. If one further checks the position of the BTPs,
one finds that when $\delta\!\approx\!0.067$\,eV,
six quadratic BTPs are created
in the mirror planes, which then split into twelve Weyl nodes
and propagate in the BZ following the solid black and
dashed red lines in Fig.~\ref{fig:6band-all}(c)
and (d). These Weyl nodes eventually annihilate with each other
at $\delta\!\approx\!0.074$\,eV after exchanging partners, which
qualitatively reproduces the phase-transition behavior
of the VCA effective Hamiltonians very well.
When $\delta\!>\!0.074$\,eV, the system becomes a strong TI.
The bulk bandstructure at $\delta\!=0.09\,$eV in the TI phase
is shown in Fig.~\ref{fig:6band-all}(a), which very well captures the
low-energy dispersions around $Z$ that were found in the first-principles
calculations.

%

\subsubsection{Discussion}

To conclude this section, we would like to comment that
the width of the WSM phase depends on two ingredients.
On one hand, as discussed above, 
it depends on the symmetry of the system; other things being equal,
the WSM interval tends to be wider in systems with lower symmetry.
On the other hand, even for fixed symmetry, it also depends on the
the detailed choice of path connecting the topological and
trivial phases. Choosing a different path may broaden or
reduce the WSM region. For example, if one artificially changes the
strength of the atomic SOC strength in \labi\ and \lubi\ in the Wannierized
TB models, and scales the variation of the actual atomic SOC by a single
scaling parameter $\lambda$, then we find that the WSM region only
shows up for $\lambda$ in the range of 76.8-77.3\%, which is
significantly narrower than for the VCA case.
However, if an average SOC is applied to the entire system, such
that the SOC strength on Te is artificially high and that on
Bi is artificially low, we find that a much wider WSM region results.
Thus, it may potentially be possible to engineer the width of a
critical WSM phase if one can modify the transformation path,
as by epitaxial strain, pressure, or additional chemical
substitution.

\subsection{BiTeI: revisited}
\label{sec:bitei}
%
\begin{figure}
\subfigure{
\includegraphics[width=6cm]{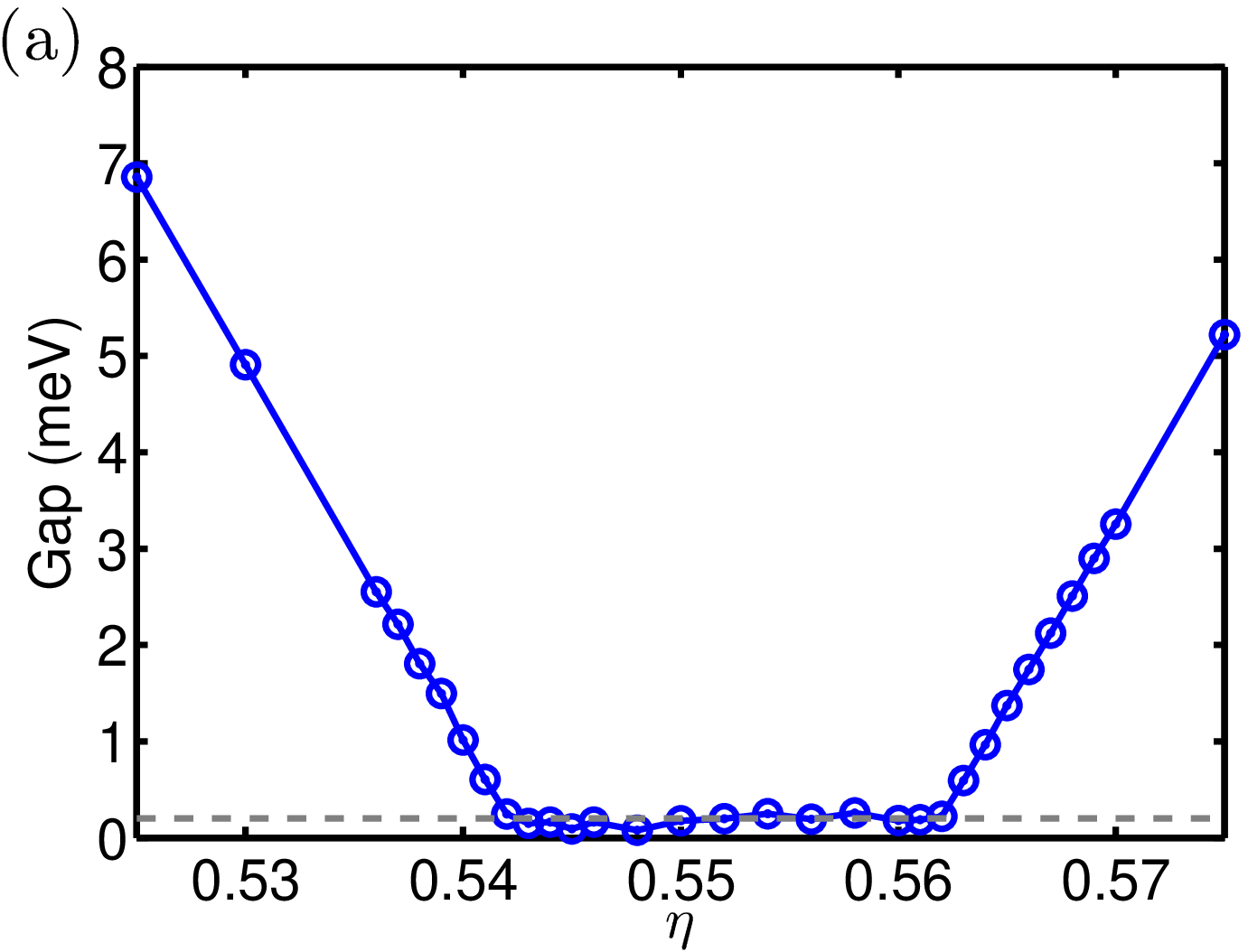}}
\subfigure{
\includegraphics[width=6.5cm]{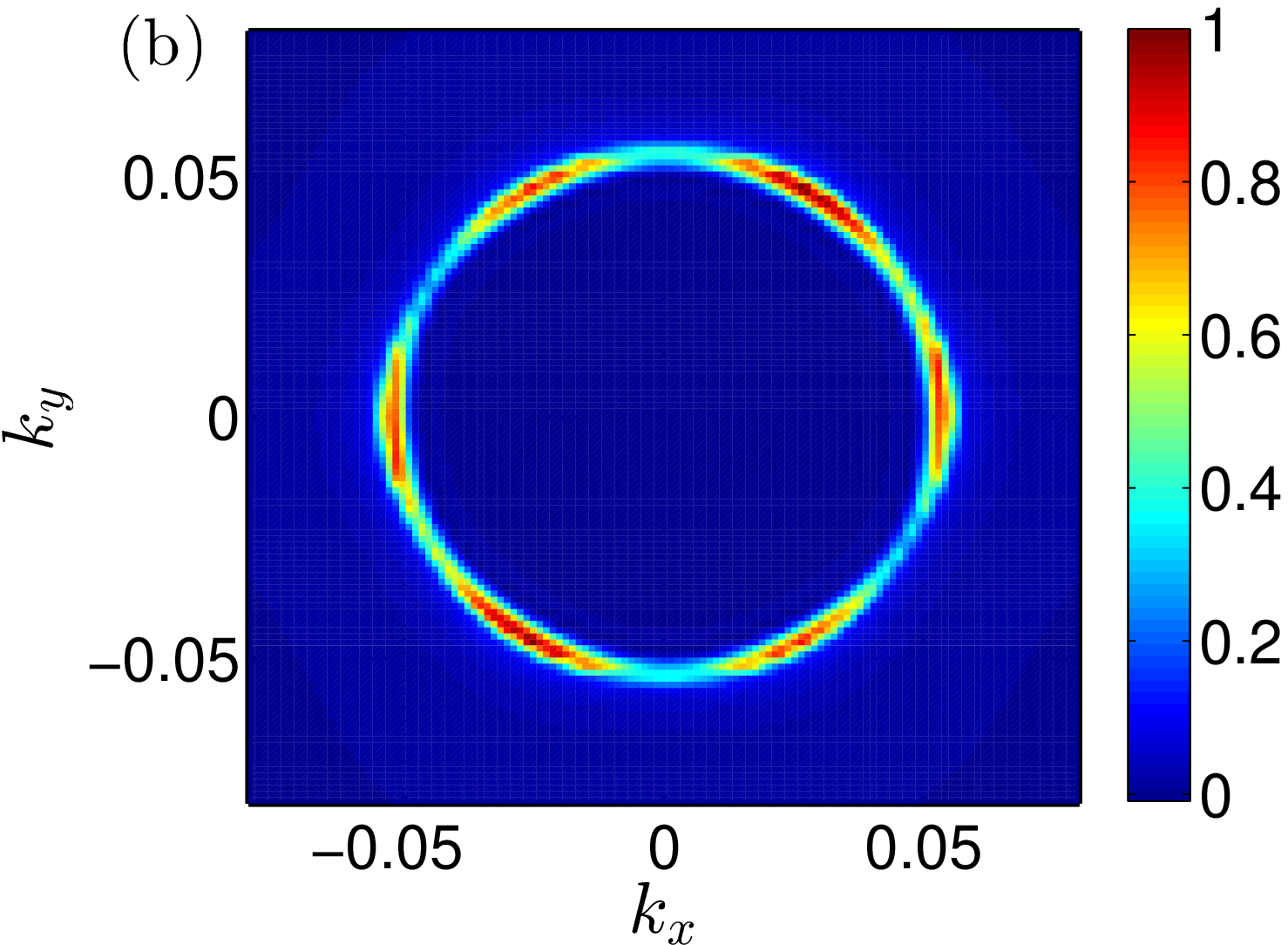}}
\caption{
(a) Smallest direct band gap in the BZ of BiTeI vs.~the
pressure-scaling variable $\eta$.
(b) Surface spectral function of BiTeI in the WSM phase
(at 55\% of the full pressure).
}
\label{fig:bitei1}
\end{figure}

In order to justify the discussion in Sec.~\ref{subsec:general-bitei},
we revisit
the TPT in BiTeI driven by pressure. In our calculations, the pressure is
applied by compressing the volume of the primitive cell.
The fully compressed volume $V$ is taken to be 85.4\% of the original
volume $V_0$, such that the former is well inside
the topological region,\cite{bahramy-naturecomm-12}
and both the lattice vectors
and atomic positions are relaxed at the compressed
volume. As discussed in Sec.~\ref{sec:prelim-method},
we searched for BTPs over the entire irreducible BZ for a transitional
Hamiltonian scaled as $H(\eta)\!=\!(1-\eta)H_0+\eta H_1$
for $0\!\le\!\eta\!\le\!1$, where
$H_0$ and $H_1$ represent the Hamiltonians of the uncompressed and
fully compressed BiTeI, with even and odd \Z2\ indices respectively.
As shown in Fig.~\ref{fig:bitei1}(a), as the pressure is increased
from 0\% to 100\% (alternatively, as $V$ is deceased from
100\% to 85.4\% of $V_0$), a semimetallic phase emerges for
$\eta$ in the range of about 54-56\%.

\begin{figure}
\subfigure{
\includegraphics[width=4.03cm]{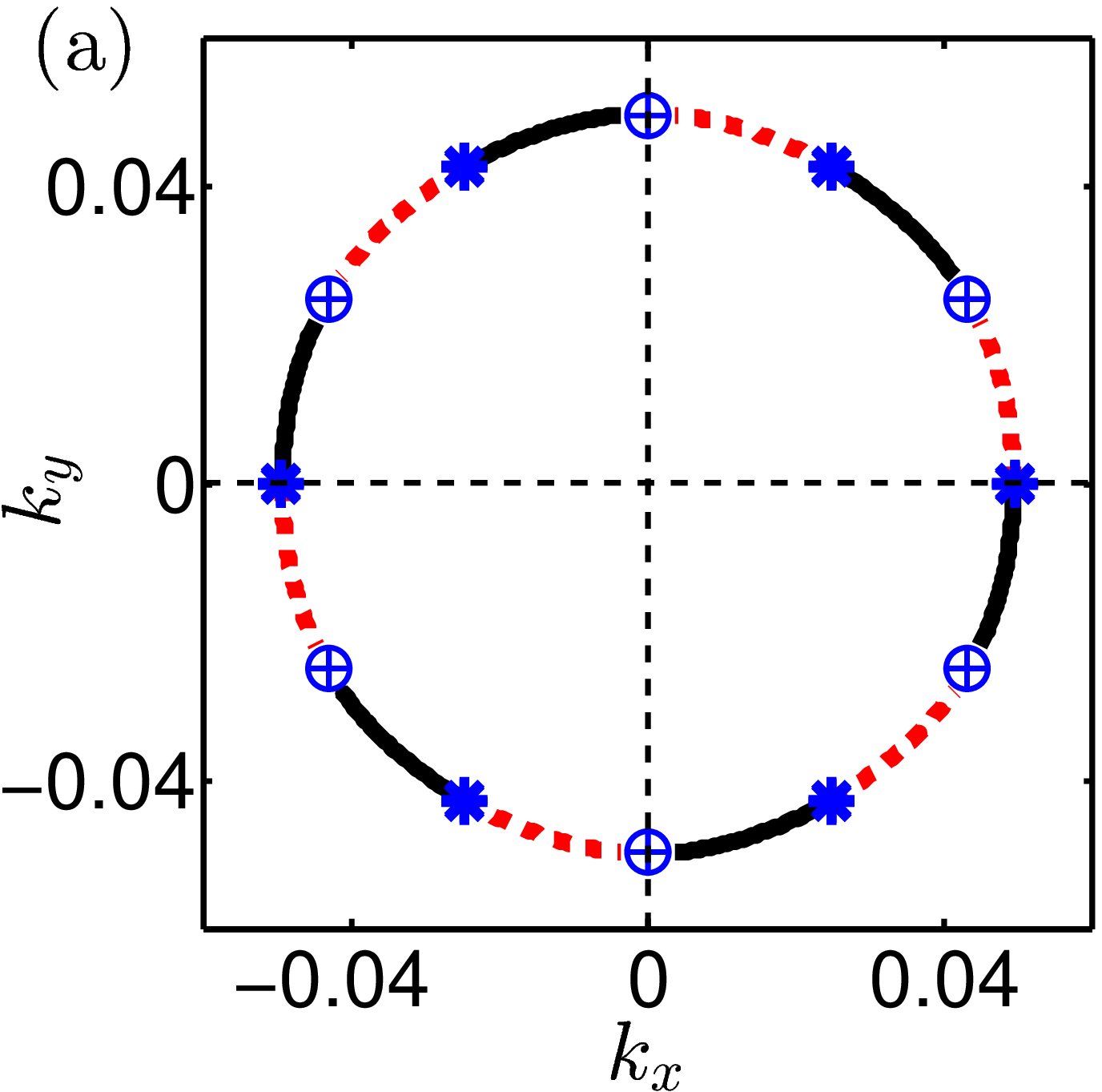}}
\subfigure{
\includegraphics[width=4.27cm]{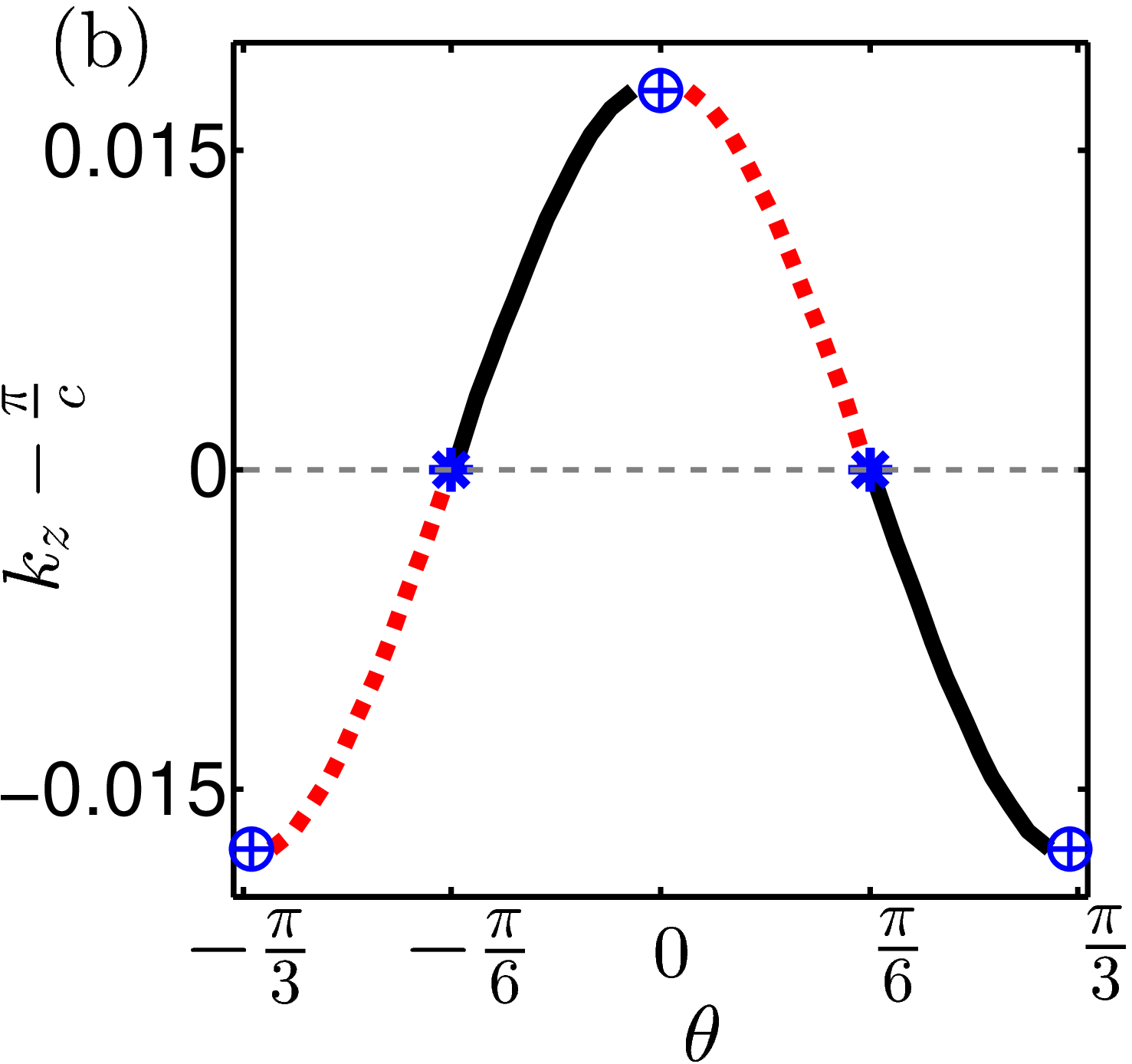}
}
\caption{(a) Trajectories of Weyl nodes in the ($k_x, k_y$) plane
(in units of \AA$^{-1}$). Dashed red (solid black) lines indicate the
trajectories of Weyl nodes with positive (negative) chirality. 
The ``*" and ``$\oplus$" denote the creation and annihilation point
of the Weyl nodes respectively.
(b) Trajectory of Weyl nodes in the $k_z$ direction (units of \AA$^{-1}$).
}
\label{fig:bitei2}
\end{figure}

The point group of BiTeI is the same as for \labi\ and \lubi,
namely C$_{3v}$.  Therefore, as explained in Sec.~\ref{sec:labisb},
one would expect the emergence of
twelve Weyl nodes in the entire BZ during the phase-transition process.
The trajectories of the Weyl nodes are plotted in
Fig.~\ref{fig:bitei2}(a-b). When $\eta\!\approx\!54\%$,
six quadratic BTPs are first created
at the BZ boundary $k_z\!=\!\pi/c$ in the $k_y\!=\!0$ and other
equivalent high-symmetry planes.
These BTPs then split into twelve Weyl
nodes which propagate along the directions indicated by solid black
(antimonopoles) and dashed red (monopoles) lines. 
They annihilate each other in the three mirror planes after 
exchanging partners. Note that in this case the system goes from a
normal to topological insulator as $\eta$ increases, which is the
reverse of the \labisb\ and \lubisb\ cases.

The results shown in Fig.~\ref{fig:bitei2} support
our conclusions in Sec.~\ref{subsec:general-bitei}.
In particular, even though the torsion argument implies that the
trajectories of the two Weyl nodes which split off from a given
quadratic BTP would never meet each other, a closed curve is still
formed in the 3D BZ of BiTeI through the interchange of partners
among the Weyl nodes.

Fig.~\ref{fig:bitei1}(b) shows the surface spectral function
of BiTeI averaged around the Fermi level for $\eta\!=\!0.55$, in the
WSM phase. It is clear that there are six
Fermi arcs extending between the six pairs of projected Weyl nodes,
which is again the hallmark of a WSM phase.

We therefore conclude that a WSM phase does exist in the TPT of BiTeI,
but it occurs only within a narrow pressure range.
If $\eta$ is changed by 2.5\%, the volume is
only changed by 0.39\%, which might be difficult to
measure experimentally. Again, the narrowness of 
the WSM interval can be attributed
in part to the high symmetry of the system.
However, as emphasized in the previous section,
the width of the critical WSM is also sensitive to the choice of
path in parameter space. The critical WSM
could get broadened by choosing a different path, as for example
by applying uniaxial
pressure. We leave this for a future study.

\section{Summary}
\label{sec:summary}

In this paper, we have investigated the nature of the TPT in a
noncentrosymmetric TI in the most general case.
We find that an intermediate WSM phase is always present,
regardless of other lattice symmetries, as long as inversion
symmetry is absent.  We discussed separately the cases in which
the Jacobian matrix is rank-one or rank-two when the gap first
closes.  In the rank-two case, each
quadratic BTP would always split into a pair of Weyl nodes, which
annihilate each other after exchanging partners.
If the rank of the Jacobian is one, then the doubly-quadratic
BTP in this case would either split into four Weyl nodes, or
else immediately be gapped out again, corresponding to an
``insulator-insulator transition." However, in the latter case,
the bulk \Z2\ indices are not expected to change.
Therefore, we conclude that \Z2-even\ and \Z2-odd\ phases of
a noncentrosymmetric insulator must always be separated by a
region of WSM phase, even if other symmetries are present.

To illustrate our conclusions,
we have carried out calculations
on specific noncentrosymmetric insulators.
For \labisb\ and \lubisb\ we have used Wannierized
VCA Hamiltonians to find a WSM phase in the region
$x\!\approx\!38.5\%-41.9\%$ and $x\!\approx\!40.5\%-45.1\%$ respectively.
A six-band TB model was also constructed to describe the
topological and critical behavior in these materials.
We found that the width of the critical WSM phase can be highly
sensitive to the choice of path in the parameter space, suggesting
that there is flexibility to engineer the WSM phase.

We have also revisited the TPT of BiTeI as a function of pressure,
where previous work suggested the absence of a WSM phase.\cite{yang-prl13}
Using a carefully constructed algorithm to search for the minimum gap
in the full three-dimensional BZ, we found that a WSM phase is indeed
present over a narrow interval of pressure, although this range may
be so narrow as to make its experimental observation difficult.

In summary, we have clarified the theory of a general \Z2-even
to \Z2-odd topological phase transition in a three-dimensional
time-reversal-invariant insulator with broken inversion symmetry,
and demonstrated that an intermediate WSM phase must always be
present.
We have also detailed the behavior of \labisb\ and \lubisb\ as
promising candidates for WSMs of this kind.
While we have not considered disorder or interactions explicitly,
we expect our conclusions to survive at least for weak disorder
or interactions.
Our work is a step forward in the general understanding of topological
phase transitions, and may provide useful guidelines for the
experimental realization of new classes of Weyl semimentals.

\acknowledgments

This work is supported by NSF Grants DMR-10-05838 and 14-08838.

\bibliography{wsm}

\end{document}